\documentclass[aps,showpacs,amsmath,amssymb,superscriptaddress,nofootinbib,twocolumn]{revtex4}
\usepackage{multirow}
\usepackage{bigstrut}
\usepackage{amssymb}
\usepackage{amsmath}
\usepackage{graphicx}
\usepackage{dcolumn}
\usepackage{bm}
\usepackage{CJK}
\usepackage{relsize}
\usepackage{cancel}
\usepackage{color}

\begin{document}

\title{Spin-orbit coupling induced two-electron relaxation in silicon donor pairs}

\author{Yang Song}\email{ysong128@umd.edu}\affiliation{Condensed Matter Theory Center and Joint Quantum Institute, Department of Physics, University of Maryland, College Park, Maryland 20742}
\author{S.~Das Sarma}
\affiliation{Condensed Matter Theory Center and Joint Quantum Institute, Department of Physics, University of Maryland, College Park, Maryland 20742}

\begin{abstract}
We unravel theoretically a key intrinsic relaxation mechanism among the low-lying singlet and triplet donor-pair states in silicon, an important element in the fast-developing field of spintronics and quantum computation. Despite the perceived weak spin-orbit coupling (SOC) in Si, we find that our discovered relaxation mechanism, combined with the electron-phonon and inter-donor interactions, drives the transitions in the two-electron states over a large range of donor coupling regime. The scaling of the relaxation rate with inter-donor exchange interaction $J$ goes from $J^5$ to $J^4$ at the low to high temperature limits. Our analytical study draws on the symmetry analysis over combined band, donor envelope and valley configurations. It uncovers naturally the dependence on the donor-alignment direction and triplet spin orientation, and especially on the dominant SOC source from donor impurities. While a magnetic field is not necessary for this relaxation, unlike in the single-donor spin relaxation, we discuss the crossover behavior with increasing Zeeman energy in order to facilitate comparison with experiments.
\end{abstract}
\maketitle

\section{Introduction}\label{sec:intro}

A pair of coupled donors in an enriched $^{28}$Si host crystal forms the most cleanly defined two-qubit system or singlet-triplet ($S$-$T$) qubit in solid state materials \cite{Zwanenburg_RMP13, Shankar_PRB15}. The nearly noise-free crystal environment \cite{Tyryshkin_PRB03, Muhonen_NatNano14, Itoh_MRS14} together with the reproducible donor properties fixed by nature \cite{Kohn_SSP57} places an essential role for the intrinsic energy relaxation among the lowest two-electron states by electron-phonon (e-ph) interaction  in studying the coherence limit of quantum computation involving such qubits \cite{Borhani_PRB10, Hsueh_PRL14}. A proper treatment of this problem with a transparent and physical basis is imperative, enabling its various modifications such as proximity with interface \cite{Calderon_PRB07}, gate potential \cite{Dehollain_PRL14, Gonzales-Zalba_NanoLett14}, or quantum dots \cite{Foote_APL15, Harvey-Collard_arxiv15} which are necessary for qubit control and operation.

Electron spin relaxation of single donor states in Si has been studied since the 1950s. Much interest then was associated with the ensemble spin resonance experiments which could remarkably map out the detailed donor wavefunction by its hyperfine interaction with the donor or $^{29}$Si nucleus \cite{Fletcher_PR54,Feher_PR59}. Theoretical attention was thus originally paid to the simultaneous flip of electron and donor nucleus spins ($T_X$) \cite{Pines_PR57} through hyperfine interaction, until later the importance of spin-orbit coupling (SOC) driven spin relaxation was established \cite{Hasegawa_PR60_Roth_PR60_Roth_LincolnLab60,Wilson_PR61, Castner_PR63, Castner_PR67}. However, the two-electron relaxation  on a coupled donor pair has been much less studied over the decades, with only a few works in the context of concentration-dependent spin relaxation \cite{Sugihara_JPCS68} or of quantum computation \cite{Borhani_PRB10}. Both of these previous studies focused on the hyperfine interaction to mix $S$ and $T$ states, while neglecting SOC as too weak in Si.

Here we unravel a distinct  \textit{general} relaxation mechanism among two-electron states to shed important light on this long-standing physical problem (i.e., electron relaxation in high density coupled donors), motivated by the development of Si donor-based spin qubits. This robust mechanism has a parametrically dominating $J^5$ dependence on the 2-donor exchange coupling $J$ for the $S\leftrightarrow T$ transition, and becomes always dominant at sufficiently strong coupling. This mechanism relies only on the intrinsic SOC in the system to couple spins, with  the donor SOC dominating the Si host SOC, but not on any structure-induced Rashba or Dresselhaus field. This particular dominance of the donor SOC is absent in the classical single-donor spin relaxation by $g$-factor modulation \cite{Hasegawa_PR60_Roth_PR60_Roth_LincolnLab60}.
The effectiveness of the two SOCs, i.e., of the host and the impurity, is considered by taking full account of the three physical `layers' in the problem: the Bloch bands, the hydrogenic-like donor envelopes, and the multi-valley configurations. Within each layer rigourous selection rules are enforced for e-ph, SOC and inter-donor perturbations, while at the same time, the decoupling between the underlying fast-oscillatory physics and the slowly varying localized envelopes realizes an efficient leading-order quantitative estimation. The unique donor and spin alignment dependence naturally falls out of our analytical treatment. Finally,  we elucidate the crossover from low to high Zeeman energy  as a reduction of the two-electron problem into a single-electron one, although the presence of spin splitting is not a priori essential for our relaxation mechanism.

The rest of the paper is organized as follows.  In Sec.~\ref{sec:derivation}, we lay out the theoretical framework, and focus on developing a suitable strategy in order to have a clear insight and efficient calculation of the two-electron state relaxation matrix elements. The general idea follows the tradition of utilizing symmetry properties to reduce the complexity as much as possible, and to proceed with the physically correct grasp (not to overlook dominant contributions or cancellations) using  perturbative approaches. To keep the flow of theoretical development, we defer several particularly long derivation details to the Appendices. In Sec.~\ref{sec:quantitative}, we make approximate quantitative evaluations for various relaxation transitions and lifetimes based on well-known physical parameters of Si band and donor energy structures, phonon dispersion, and deformation potentials. We also analyze the relaxation trend in two-electron states with external magnetic field going from zero to large values.  A summary is given in Sec.~\ref{sec:summary}.

\section{Theoretical derivation}\label{sec:derivation}

To properly treat the relaxation problem driven by SOC in a combined system of Si host and coupled double donors, the potential energy part of the Hamiltonian ($V$) for the donor outer electrons includes both the Si crystal potential ($V_{\rm cryst}$) and the difference between Si atoms and the substitutional donor ions ($\delta V_{{\rm ion},j}$, where $j=\alpha,\beta$ labels the donors). Combined with the kinetic energy ($K$) and electron-electron interaction ($V_{ee}$), the Hamiltonian for the electron states in the absence of e-ph interaction is,
\begin{eqnarray}
H_{ee} = \sum_{i=1,2} [K(\mathbf{r}_i)  + V(\mathbf{r}_i)] + V_{ee}(\mathbf{r}_1, \mathbf{r}_2),\qquad \label{eq:H_ee}
\\
K= -\frac{\hbar^2\nabla^2}{2m},\; V= V_{\rm cryst} +\sum_{j=\alpha,\beta} \delta V_{{\rm ion},j}, \;
 V_{ee} =  \frac{e^2}{\epsilon r_{12}},\label{eq:K_V_Vee}
\end{eqnarray}
where $r_{12}\equiv |\mathbf{r}_1-\mathbf{r}_2|$, $\epsilon$ is the dielectric constant in Si,  $V_{\rm cryst}=V_{{\rm cryst},0}+ V_{\rm cryst,so}$ contains spin-independent and SOC parts, and $\delta V_{{\rm ion},j}(\mathbf{r}) = -e^2/\epsilon |\mathbf{r}-\mathbf{R}_j| +\delta V_{\rm ion, cc} (\mathbf{r}-\mathbf{R}_j)$ contains the Coulomb potential as well as the short-range central-cell correction at the core of the ion. $\delta V_{\rm ion, cc} =\delta V_{\rm ion, cc,0} +\delta V_{\rm ion, cc, so}$ and $\delta V_{\rm ion, cc, so}$ easily dominates the donor SOC effect due to the fast changing core potential \cite{Weisz_PR66}, compared with that from the smooth Coulomb potential away from the central cell region. The e-ph interaction that induces the electronic transition has the general form within the harmonic approximation,
\begin{eqnarray}
H_{\rm ep} = -\sum_{k} \delta \mathbf{R}_k \cdot \bm\nabla V(\mathbf{r}-\mathbf{R}_k),\label{eq:H_ep}
\end{eqnarray}
where the lattice displacement at atom $k$ is a superposition of phonon states, $ \delta \mathbf{R}_k = \sum_\mathbf{q} \sqrt{\hbar/2 \omega^\lambda_\mathbf{q} M} (a^\lambda_\mathbf{q} \bm\xi^\lambda_{k,\mathbf{q}} e^{i\mathbf{q}\cdot \mathbf{R}_k})+$H.c, with $M$ being the mass of Si crystal, $\omega^\lambda_\mathbf{q}$, $a^\lambda_\mathbf{q}$ and $\bm\xi^\lambda_{k,\mathbf{q}}$ the eigen-frequency of the $\lambda$th phonon band at wavevector $\mathbf{q}$, the associated annihilation operator and the normalized phonon polarization vector, respectively \cite{Maradudin_SSP63}. Here we use the long-wavelength acoustic phonons in bulk Si which can compensate the small energy difference ($\lesssim$ meV) in the two-electron states. As we will see, these two-electron relaxations require the third-order perturbation theory even using the bound Bloch states as basis. As a result, the relevant intermediate states encompass both large and small energy scales due to the various fast (SOC and e-ph) and slow (exchange) perturbative terms. We are forced to go beyond the usual conductance-band-bottom effective mass approximation \cite{Kohn_SSP57} to include other Bloch bands, in addition to the central cell correction we have explicitly taken into account. This study thus represents a complicated single-particle problem which combines crystal and atomic physics along with a relatively simple many-body (two-body) problem. It is important to choose a systematic treatment for efficient calculations and avoiding pitfalls.

We start with a set of two-electron representations satisfying the symmetry of the specific two-donor system. While these states mainly comprise the Heitler-London  states out of unperturbed single-donor ground states, both nonpolar and polar (`ionic') mixtures are included. These excited components are needed for incorporating nonzero e-ph and SOC coupling, and are mixed in by the inter-donor interaction. The dominant effect comes from the leading-order mixture where one of the two electrons occupies the single-donor ground state. As a result, the general states follow the concise expressions,
\begin{eqnarray}
S&:& \; \frac{ 1-\mathcal{P}_{12}}{2\sqrt{1 + \chi^2}} [\psi^S_{\alpha\uparrow}(\mathbf{r}_1)\psi^S_{\beta\downarrow}(\mathbf{r}_2)-\psi^S_{\alpha\downarrow} (\mathbf{r}_1) \psi^S_{\beta\uparrow}(\mathbf{r}_2)]  , \quad\label{eq:S}
\\
T_0&:& \; \frac{ 1-\mathcal{P}_{12}} {2\sqrt{1 - \chi^2}}
[\psi^{T_0}_{\alpha\uparrow} (\mathbf{r}_1) \psi^{T_0}_{\beta\downarrow}(\mathbf{r}_2)+ \psi^{T_0}_{\alpha\downarrow} (\mathbf{r}_1) \psi^{T_0}_{\beta\uparrow}(\mathbf{r}_2)],\label{eq:T0}
\\
T_{\pm}&:& \; \frac{1-\mathcal{P}_{12}} {\sqrt{2(1 - \chi^2)}}
[ \psi^{T_{+(-)}}_{\alpha\uparrow(\downarrow)} (\mathbf{r}_1) \psi^{T_{+(-)}}_{\beta\uparrow(\downarrow)}(\mathbf{r}_2)], \label{eq:T+-}
\end{eqnarray}
where  $\mathcal{P}_{12}$ exchanges electrons 1 and 2,  the one-electron wavefunction $\psi_{\alpha(\beta)}$ is \textit{mainly} located around donor  $\alpha (\beta)$, $\chi = \langle \psi_\alpha |\psi_\beta\rangle$, superscripts ($S, T_{0,\pm}$)  distinguish different mixture components which are described in details later [Eq.~(\ref{eq:001_Td_mix-alphaUp})], and finally spins $\uparrow$ and $\downarrow$ (which generally are quasi-spins that contain small opposite spin component due to SOC-induced mixing) are along the $\alpha-\beta$ donor-alignment direction ($\hat{\mathbf{d}}$), dictated by symmetry.

We next classify the two-electron states and the e-ph interaction by the total system symmetries for three crystallographic directions $\hat{\mathbf{d}}$, listed in Table~\ref{tab:IR}.  For brevity, we use strain element $\epsilon_{ij}$ in place of the e-ph interaction $\sim[r_i dV/dr_j+(i\leftrightarrow j)]/2$ which transform the same way symmetry-wise.  For $\hat{\mathbf{d}}\|$[110], $S$ or $T_-+T_+$ denotes the eigenstate with the \textit{majority} $S$ or $T_-+T_+$ component respectively since they belong to the same irreducible representation (IR) .

\begin{table}[h]
\caption{\label{tab:IR}
For each $\hat{\mathbf{d}}$, [001], [111] or [110], we find its point group and  the IRs of the states and phonon modes. For $\hat{\mathbf{d}}\|$[111], two subcases, with inversion symmetry ($i$) and without it ($n$), depend on the alternating donor positions (even though individual donors possess no inversion symmetry). To be definite, we set $z'\|[111]$, $x'\| [10-1]$, $y'\|[-1 2-1]$,  $z''\|[110]$, $x''\|[001]$ and $y''\|[1-10]$.
}
\renewcommand{\arraystretch}{1.5}
\tabcolsep=0.1 cm
\begin{tabular}{cccc}
 \hline\hline
            &  [001] $(D_{2d})$   & $ \renewcommand{\arraystretch}{1.0}\begin{array}{c} \![111]n\; (C_{3v}) \\ \![111]i\; (D_{3d})\end{array}$ & [110] $(C_{2v})$ \\
               \hline
$S$ &   $A_1$ & $A_1$ & $A_1$\\
$T_0$ &  $B_1$ & $A_2(n), A'_1(i)$ & $A_2$\\
$T_\pm$ & $E$ & $E(n),  E'(i)$ & $ \renewcommand{\arraystretch}{1.0}\begin{array}{c} A_1(T_++T_-) \\ B_2 (T_+-T_-) \end{array}$ \\
       & $A_1\left( \!\!\renewcommand{\arraystretch}{0.8}\begin{array}{c} \epsilon_{zz} \\ \epsilon_{xx}+\epsilon_{yy} \end{array}\!\!\right)$ & $A_1\left( \!\!\renewcommand{\arraystretch}{0.8}\begin{array}{c} \epsilon_{z'z'} \\ \epsilon_{x'x'}+\epsilon_{y'y'} \end{array}\!\!\right)$ & $A_1\left( \!\!\renewcommand{\arraystretch}{0.8}\begin{array}{c} \epsilon_{x''x''} \\ \epsilon_{y''y''} \\ \epsilon_{y''y''} \end{array}\!\!\right)$
\\
$ \renewcommand{\arraystretch}{0.8}\begin{array}{c} \epsilon_{ij} \\  \\ \\\end{array}$     & $ \renewcommand{\arraystretch}{1.0}\begin{array}{c} B_1 (\epsilon_{xx}-\epsilon_{yy} ) \\ B_2 (\epsilon_{xy}) \\ E(\{\epsilon_{xz},\epsilon_{yz}\} ) \end{array}$ &
     $ E \left(\renewcommand{\arraystretch}{0.8}\begin{array}{c} \!\! \{\epsilon_{x'x'}-\epsilon_{y'y'} ,\\ 2\epsilon_{x'y'}\} \\ \{\epsilon_{x'z'},\epsilon_{y'z'}\} \!\! \end{array} \!\!\right)$ &
     $ \renewcommand{\arraystretch}{0.8}\begin{array}{c} B_1 (\epsilon_{x''z''}) \\ B_2 (\epsilon_{x''y''}) \\ A_2 (\epsilon_{y''z''} ) \end{array}$
\\
\hline\hline
\end{tabular}
\end{table}

This top-down symmetry approach immediately identifies the allowed relaxation processes and the associated specific acoustic phonon modes, independent of the quantitative treatments one adopts. The nonvanishing phonon-induced relaxation matrix elements are,
\begin{eqnarray}
&\langle T_0| \epsilon_{xx}-\epsilon_{yy} | S \rangle_{[001]} , \; \langle T_\pm | \epsilon_{xz}\mp (\pm) i\epsilon_{yz} | S (T_0)\rangle_{[001]} ,&
 \label{eq:001_coupling}
\\
 &\langle T_\pm | \epsilon_{x'z'}\pm i \epsilon_{y'z'} , \epsilon_{x'x'}-\epsilon_{y'y'} \pm 2i
\epsilon_{x'y'}  | S,  T_0  \rangle_{[111]n} ,&
\label{eq:111n_coupling}
\\
 & \langle T_\pm | \epsilon_{x'z'}\pm i \epsilon_{y'z'}, \epsilon_{x'x'}-\epsilon_{y'y'}\pm 2i
\epsilon_{x'y'}| T_0 \rangle_{[111]i} ,
& \label{eq:111i_coupling}
\\
& \langle T_0 |\epsilon_{y''z''} | S ,    T_1 \rangle_{[110]}  ,\; \langle T_2 | \epsilon_{x''z''} | T_0 \rangle_{[110]}, &
 \label{eq:110_coupling_1}
\\
&
 \langle T_1 | \epsilon_{x''x''}, \epsilon_{z''z''}, \epsilon_{y''y''}| S \rangle_{[110]} ,\; \langle T_2 | \epsilon_{x''y''}| S, T_1 \rangle_{[110]} .& \quad
 \label{eq:110_coupling}
\end{eqnarray}
where $T_{1/2}\equiv T_+\pm T_-$ in Eqs.~(\ref{eq:110_coupling_1}) and (\ref{eq:110_coupling}). We also have taken into account the fact that $S$ and $T$ states (e-ph operator) are two(one)-body objects, removing all $\langle T_-| \epsilon_{ij}| T_+ \rangle$ where $\langle \psi^{T_-}_{\alpha(\beta)\downarrow}| \psi^{T_+}_{\alpha\uparrow}\rangle$ are always strictly zero, solely by the $C_{2(3)}$ operation in the $\hat{\mathbf{d}}\|$[001]([111]) case.

To evaluate the allowed relaxation matrix elements, the first simplifications come from substituting the forms of $S$ and $T$ states with Eqs.~(\ref{eq:S})-(\ref{eq:T+-}). Even though the specific $\psi$'s are altered away from the single-donor wavefunction by inter-donor interaction, they still obey precise symmetry relations among themselves due to the point group of a given two-donor system. Utilizing these relations, the matrix elements are reduced into products of single-electron ones. A representative set of steps is shown for $\langle T_0| \epsilon_{xx}-\epsilon_{yy} | S \rangle_{[001]}$,
\begin{eqnarray}
&&\langle T_0| \epsilon_{xx}-\epsilon_{yy} | S \rangle_{[001]}
\nonumber\\
&\propto&\!\!  \frac{1}{4} \langle \psi^{T_0}_{\alpha\uparrow}(\mathbf{r}_1) \psi^{T_0}_{\beta\downarrow}(\mathbf{r}_2) \!+ \! \psi^{T_0}_{\alpha\downarrow}(\mathbf{r}_1) \psi^{T_0}_{\beta\uparrow}(\mathbf{r}_2)
\! -\!(1 \leftrightarrow 2) |
\nonumber\\
&&\quad[\epsilon_{xx}(\mathbf{r}_1)\!-\!\epsilon_{yy} (\mathbf{r}_1)]
\nonumber\\
&&\quad| \psi^{S}_{\alpha\uparrow}(\mathbf{r}_1) \psi^{S}_{\beta\downarrow}(\mathbf{r}_2) \!- \! \psi^{S}_{\alpha\downarrow}(\mathbf{r}_1) \psi^{S}_{\beta\uparrow}(\mathbf{r}_2)
\! -\!(1 \leftrightarrow 2)  \rangle \quad
\nonumber\\
&=&\frac{1}{4} \sum_{\substack {\updownarrow=\uparrow,\downarrow \\ \gamma=\alpha,\beta}}
(-1)^{\updownarrow+\gamma}
\left\{\langle \psi^{T_0}_{\gamma\updownarrow}|\epsilon_{xx}-\epsilon_{yy}| \psi^{S}_{\gamma\updownarrow}\rangle \langle \psi^{T_0}_{\overline{\gamma}\overline{\updownarrow}} | \psi^{S}_{\overline{\gamma}\overline{\updownarrow}} \rangle
\right.
\nonumber\\
&& \qquad \qquad\left.-
\langle \psi^{T_0}_{\gamma\updownarrow}|\epsilon_{xx}-\epsilon_{yy}| \psi^{S}_{\gamma\overline{\updownarrow}}\rangle \langle \psi^{T_0}_{\overline{\gamma}\overline{\updownarrow}} | \psi^{S}_{\overline{\gamma}\updownarrow} \rangle
\right.
\nonumber\\
&&  \qquad \qquad \left.+
\langle \psi^{T_0}_{\gamma\updownarrow}|\epsilon_{xx}-\epsilon_{yy}| \psi^{S}_{\overline{\gamma}\updownarrow}\rangle \langle
\psi^{T_0}_{\overline{\gamma}\overline{\updownarrow}} | \psi^{S}_{\gamma\overline{\updownarrow}} \rangle
\right.
\nonumber\\
&&  \qquad \qquad \left.
-\langle \psi^{T_0}_{\gamma\updownarrow}|\epsilon_{xx}-\epsilon_{yy}| \psi^{S}_{\overline{\gamma}\overline{\updownarrow}}\rangle \langle
\psi^{T_0}_{\overline{\gamma}\overline{\updownarrow}} | \psi^{S}_{\gamma\updownarrow} \rangle
\right\}
\nonumber\\
&=& \! \! \langle \psi^{T_0}_{\alpha\uparrow}\!|\epsilon_{xx}\!-\!\epsilon_{yy}| \psi^{S}_{\alpha\uparrow}\!\rangle \langle \psi^{T_0}_{\beta\downarrow} \!| \psi^{S}_{\beta\downarrow} \rangle \!
\textrm{\! (by \!$\sigma_{[110]}$,\! $C_{2x}$,\! $C_{2z}$)}
\nonumber\\
&&\! \!-\langle \psi^{T_0}_{\beta\uparrow}\!|\epsilon_{xx}\!-\!\epsilon_{yy}| \psi\!^{S}_{\alpha\uparrow}\rangle \langle
\psi\!^{T_0}_{\alpha\downarrow} | \psi\!^{S}_{\beta\downarrow} \!\rangle\!\!
\textrm{ (by \!$\sigma_{[110]}$, \!\!$S_{4z}$, \!\!$C_{2z}$)}.\quad \label{eq:S-T0-001}
\end{eqnarray}
where for brevity we  use `$\propto$' and omit the normalization factor $1/\sqrt{1\pm \chi^2}$, $\overline\gamma$ or $\overline{\updownarrow}$ denotes the opposite donor or spin respectively, and at exponent $\alpha,\uparrow\equiv0$ and $\beta,\downarrow\equiv1$. The key symmetry operations used for the reduction are explicitly marked in the parenthesis. Similarly, we can substitute the expressions from Eqs.~(\ref{eq:S})-(\ref{eq:T+-}) in the rest of relaxation matrix elements in Eqs. (\ref{eq:001_coupling})-(\ref{eq:110_coupling}), and utilize available symmetry operations to make the simplifications.

\begin{widetext}
\begin{eqnarray}
&\langle T_0|\epsilon_{y''z''} | S \rangle_{[110]}
\propto
\sum_{\substack {\updownarrow=\uparrow,\downarrow \\ \gamma=\alpha,\beta}}
(-1)^{\gamma}\langle \psi^{T_0}_{\gamma\updownarrow''}|\epsilon_{y''z''}| \psi^{S}_{\alpha\uparrow''}\rangle \langle \psi^{T_0}_{\overline{\gamma}\overline{\updownarrow}''} | \psi^{S}_{\beta\downarrow''} \rangle
\textrm{ (by $\sigma_{y''}$, $C_{2x''}$ and $\sigma_{z''}$)},&\label{eq:S-T0-110}
\\
&\langle T_+| \epsilon_{xz}\mp i\epsilon_{yz} | S/T_0 \rangle_{[001]}
\propto \frac{\mp1}{\sqrt{2}}\sum_{\gamma=\alpha,\beta} (-1)^{\gamma}\langle \psi^{T_+}_{\gamma\uparrow}| \epsilon_{xz}\mp i\epsilon_{yz}| \psi^{S/T_0}_{\alpha\downarrow}\rangle \langle \psi^{T_+}_{\overline\gamma\uparrow} | \psi^{S/T_0}_{\beta\uparrow} \rangle
\textrm{ (by $S_{4z}$ and $C_{2z}$)},&\label{eq:S-T+-001}
\\
&\langle T_+| \epsilon_{x'z'}+i\epsilon_{y'z'} | S/T_0 \rangle_{[111]n}
\propto 
 \frac{\mp1}{2\sqrt{2}}\sum_{\gamma,\gamma'=\alpha, \!\beta} (-1)^{\gamma+\frac{\gamma'}{2}\mp\frac{\gamma'}{2}}
 \langle \psi^{T_+}_{\gamma\uparrow'}|  \epsilon_{x'z'}\!+\! i\epsilon_{y'z'}| \psi^{S/T_{0}}_{\gamma'\downarrow'}\rangle \langle \psi^{T_{+}}_{\overline{\gamma}\uparrow'} | \psi^{S/T_{0}}_{\!\overline{\gamma'}\uparrow'} \rangle
 \textrm{(by $C_{3z'}$)},& \label{eq:S-T+-111n}
\\
&\langle T_+| \epsilon_{x'z'}+i\epsilon_{y'z'} | T_0 \rangle_{[111]i}
\propto
 \frac{1}{\sqrt{2}} \sum_{\gamma=\alpha,\beta} (-1)^\gamma \langle \psi^{T_+}_{\gamma\uparrow'}| \epsilon_{x'z'}+i\epsilon_{y'z'}| \psi^{T_0}_{\alpha\downarrow'}\rangle \langle \psi^{T_+}_{\overline\gamma\uparrow'} | \psi^{T_0}_{\beta\uparrow'} \rangle
 \textrm{ (by $C_{3z'}$ and $i$)},&\label{eq:T0-T+-111i}
\\
&\langle T_+\pm T_-| \epsilon_{A_1}/\epsilon_{x''y''} | S \rangle_{[110]}
\propto
 \sum_{\substack {\updownarrow=\uparrow,\downarrow \\ \gamma=\alpha,\beta}} (-1)^{\updownarrow+\gamma} \langle \psi^{T_{1/2}}_{\gamma\uparrow''}| \epsilon_{A_1/x''y''}| \psi^{S}_{\alpha\updownarrow''}\rangle \langle \psi^{T_{1/2}}_{\overline\gamma\uparrow''} | \psi^{S}_{\beta\overline{\updownarrow}''} \rangle
\textrm{ (by $\sigma_{z''}$ and $\sigma_{y''}$)} ,&\label{eq:S-T12-110}
\\
&\langle T_+\pm T_-| \epsilon_{y''z''}/\epsilon_{x''z''} | T_0 \rangle_{[110]}
\propto
 \sum_{\substack {\updownarrow=\uparrow,\downarrow \\ \gamma=\alpha,\beta}} (-1)^\gamma
 \langle \psi^{T_{1/2}}_{\gamma\uparrow''}| \epsilon_{y''z''/x''z''}| \psi^{T_0}_{\alpha\updownarrow''}\rangle \langle \psi^{T_{1/2}}_{\overline\gamma\uparrow''} | \psi^{T_0}_{\beta\overline{\updownarrow}''} \rangle
\textrm{ (by $\sigma_{z''}$ and $\sigma_{y''}$)},& \label{eq:T0-T12-110}
\\
&\langle T_+ -T_-| \epsilon_{x''y''} | T_++T_- \rangle_{[110]}
\propto
\sum_{\substack {\updownarrow=\uparrow,\downarrow \\ \gamma=\alpha,\beta}} (-1)^{\gamma+\updownarrow}
\langle \psi^{T_2}_{\gamma\updownarrow''}|\epsilon_{x''y''}| \psi^{T_1}_{\alpha\uparrow''}\rangle \langle \psi^{T_2}_{\overline\gamma\updownarrow''} | \psi^{T_1}_{\beta\uparrow''} \rangle
\textrm{ (by $\sigma_{z''}$ and $\sigma_{y''}$)},& \label{eq:T-barT-110}
\end{eqnarray}
\end{widetext}
where   $\uparrow'(\uparrow'')$ indicates that the spin is along the $z'(z'')$ rather than $z$ direction, and again $T_{1/2}\equiv T_+\pm T_-$ in Eqs.~(\ref{eq:S-T12-110})-(\ref{eq:T-barT-110}). The  transition matrix element magnitudes from $S$ or $T_0$ to $T_-$ state are the same as those to $T_+$ as obtained by switching between $\epsilon_{xz}\!\pm \!i\epsilon_{yz}$ ($\hat{\mathbf{d}}\|[001]$) or $\epsilon_{x'z'}\!\pm\! i\epsilon_{y'z'}$ ($\hat{\mathbf{d}}\|[111]$). In $\hat{\mathbf{d}}\|[111]$, one can just substitute $ \epsilon_{x'z'}\!\pm\! i\epsilon_{y'z'}$ with $ \epsilon_{x'x'}- \epsilon_{y'y'}\pm 2i  \epsilon_{x'y'}$. We note that the additional inversion in the $[111]i$ case equates each of the two pairs in Eq.~(\ref{eq:S-T+-111n}) of $[111]n$, canceling  $\langle T_\pm| \epsilon_{x'z'}\!\!\pm\! i\epsilon_{y'z'}\! |\! S \rangle_{\![111]i}$ while leading to Eq.~(\ref{eq:T0-T+-111i}). This set of equations arising from very general symmetry considerations constitutes one of the key results in this work.

We apply the perturbation theory to quantify the single-electron matrix elements in terms of  SOC  and exchange coupling constants, in addition to the deformation potential coupling (i.e., e-ph interaction with acoustic phonons). The first two perturbations are necessary, as without them the transitions reduce to those between pure opposite spins or single-donor spin relaxation which must vanish. In particular, the same-spin e-ph matrix element in Eq.~(\ref{eq:S-T0-001}) from $\langle T_0| \epsilon_{xx}-\epsilon_{yy} | S \rangle_{[001]}$  requires a $z$-component SOC operator, since
\begin{eqnarray}
\langle \psi^{T_0}_{\alpha(\beta)\uparrow}|\epsilon_{xx}-\epsilon_{yy}| \psi^{S}_{\alpha\uparrow}\rangle = -\langle \psi^{T_0}_{\alpha(\beta)\downarrow}|\epsilon_{xx}-\epsilon_{yy}| \psi^{S}_{\alpha\downarrow}\rangle
\label{eq:SOCz_needed}
\end{eqnarray}
by the $\sigma_{[110]}$ reflection symmetry. It is similar for the same-spin transitions of $\hat{\mathbf{d}}\|[110]$ in Eqs.~(\ref{eq:S-T0-110}), (\ref{eq:S-T12-110})-(\ref{eq:T-barT-110}) except with the $\epsilon_{A_1}$ or $\epsilon_{x''z''}$ modes, while all the rest manifestly require SOC to flip the spin.

To proceed within the perturbation theory, we choose our basis states to be the spinless donor states, i.e., the eigenstates of $H_{e,0} = K+V_{\rm cryst,0}- e^2/\epsilon r$,
\begin{eqnarray}
\{\psi_k\} = \sum_{i=1}^6 v_{k,i} F^{nlm}_{k,i} (\mathbf{r}) \psi^{\Delta_j}_{k,i}(\mathbf{r}),
\end{eqnarray}
which are identified by three indices: the Si bulk band ($\Delta_1, \Delta_{2'}, \Delta_2, \Delta_{1'}$ and $\Delta_5$) \cite{Lax_PR61}, the donor envelope with an orbital number $nlm$ (with ellipsoidal effective mass) \cite{Kohn_PR55}, and the $T_d$ (tetrahedral) group IR ($A_1, A_2, E, T_1$ and $T_2$) \cite{Bradley_Cracknell72} which determines $v_{k,i}$ ($\sum_i |v_{k,i}|^2=1$) considering the participating $\Delta_j$ and $nlm$. These three indices fix the energy level in a roughly descending order. The three perturbations in addition to $H_{e,0}(\mathbf{r}_1-\mathbf{r}_\alpha) +H_{e,0}(\mathbf{r}_2-\mathbf{r}_\beta)$ can now be explicitly seen in $H_{ee}+H_{\rm ep}$ [from Eqs.~(\ref{eq:H_ee}) and (\ref{eq:H_ep})]: the SOC part, $V_{\rm crysta, so}$; the obvious e-ph part, $H_{\rm ep}$; and the inter-donor part, $H_{\rm int-d}=V_{ee} + \delta V_{\rm ion, \alpha}(\mathbf{r}_2) +\delta V_{\rm ion, \beta}(\mathbf{r}_1)$, which is a two-body interaction and its matrix element can be integrated over one variable (e.g., $\mathbf{r}_2$) to obtain the mixture components it contributes to a single electron wavefunction [e.g., $\psi(\mathbf{r}_1)$ in Eqs.~(\ref{eq:S})-(\ref{eq:T+-})]. The effective single-electron interaction from the inter-donor coupling, defined in this way, is denoted $\overline{H}_{\rm int-d}$, as used in Eq.~(23). Before exhaustively working out all possible selection rules among this multitude of states, we examine the essential physics of coupling strengths for different perturbation interactions and select the stronger couplings efficiently.

First, we focus on the inter-donor interaction, including direct Coulomb and exchange terms. It couples donor states made of different bulk $\Delta$ bands very weakly. As we know, in single donor ground states, bands other than the conduction $\Delta_1$ band are routinely neglected due to their fast-oscillating difference and the slowly varying nature of the Coulomb interaction \cite{Luttinger_PR55,Kohn_SSP57}. Here the coupling by inter-donor interaction is even weaker as the inter-donor distance is several times the Bohr radius. Within the same bulk band, it can couple different donor envelopes as well as valley configurations effectively, as their differences are  (partly) slowly varying.
Second, the e-ph coupling between the same or different $\Delta_j$'s  are efficient when  allowed by symmetry, as the interaction involves periodic ion potentials. This gives rise to various intraband and interband deformation potentials. Once the interaction matches the symmetry difference of the two bulk bands, it can only couple the same envelopes as  no extra symmetry from the phonon mode compensates for the different envelope symmetries. However, it may couple different valley configurations as the intravalley e-ph coupling may change from valley to valley.
Last, we discuss the SOC of two different types arising from the host and the donor \cite{footnote_2SOC}. For Si host SOC, it couples symmetry-allowed $\Delta$ bands strongly but different envelopes negligibly just like the e-ph coupling. The donor impurity SOC  may couple donor envelopes in the same band effectively \cite{Sugihara_JPSJ63, Shimizu_JPSJ70, Castner_PR67}. Additionally, both SOCs can couple different valley configurations allowed by symmetry. However, the host SOC together with the e-ph coupling connecting two same $\Delta$ bands (imposed by the inter-donor coupling) largely recovers the ``Elliott-Yafet'' (E-Y) cancellation that occurs in the bulk Si spin-phonon interaction  \cite{Yafet_SSP63, Song_PRB12}, and suppresses their effect by a large factor of the relevant phonon wavelength divided by lattice constant. Without this suppression, the spin mixing caused by the host SOC would be about $\Delta^{\rm hst}_{\rm SOC}/\Delta \mathcal{E}^{\rm hst}\sim$ 40 meV/ 4eV (interband coupling) and comparable to that by the P donor SOC, $\Delta^{\rm dnr}_{\rm SOC}/\Delta \mathcal{E}^{\rm dnr} \sim$ 0.03 meV/12 meV (inter-bound-state coupling). For more details see Eq.~(\ref{eq:S-T0-001-1}). The spin splitting in the donor states, on the other hand, depends strongly on the donor types (P, As, Sb) and comes largely from the donor SOC \cite{Aggarwal_PR65, Castner_PR67}. It measures the SOC contribution from the $T_d$ potential deviating from a spherical one, i.e, the SOC contribution that breaks the inversion symmetry and hence annuls the first-order-in-wavevector E-Y cancellation \cite{Yafet_SSP63, Song_PRB12}.


Understanding the potentially dominant couplings, one can identify the symmetry-allowed ones among them. The e-ph and SOC couplings follow conventional single-particle selection rules, as discussed below. The two-body inter-donor interaction, however, requires a different treatment. We identify the allowed $T_d$ IRs for the  mixed  single-donor states,  such that under every two-donor symmetry operation the resulting $\psi_{\gamma \updownarrow}$ transforms the same way as that for the ground state. Symmetry-wise there are totally 10 components under the $T_d$ group: $A_1, A_2, E, T_1$ and $T_2$ IRs (here we do not include SOC in this mixture, as we intend to account for inter-donor and SOC effects separately). By checking the character tables of $D_{2d}$, $C_{3v}(D_{3d})$ or $C_{2v}$ groups for the symmetry operations with $\hat{\mathbf{d}}\|$ [001], $[111]n(i)$ or [110] respectively, one can obtain the allowed mixture components by comparison.  To not get distracted from the development of the central physical idea, we list all the results systematically in Appendix~\ref{app:aaa}, providing the explicit operation matrices for clarity.
For instance, $\psi_{\alpha\uparrow}$ in Eqs.~(\ref{eq:S})-(\ref{eq:T+-}) for $\hat{\mathbf{d}}\|[001]$ follows,
\begin{eqnarray}\label{eq:001_Td_mix-alphaUp}
\alpha_\uparrow &=& \alpha_{A_1\uparrow} +\delta^\beta_0\beta_{A_1\uparrow}
+ \sum_{\gamma=\alpha,\beta}(i\delta^\gamma_1 \gamma_{A_2\uparrow}
\nonumber\\
&& + i\delta^\gamma_2 \gamma_{E^{I}_z\uparrow} + \delta^\gamma_3 \gamma_{E^{I\!I}_z\uparrow} + \delta^\gamma_4 \gamma_{T_{1z}\uparrow} + i\delta^\gamma_5 \gamma_{T_{2z}\uparrow}),\quad
\end{eqnarray}
where `$\psi$' is omitted for shortness, and both the nonpolar ($\gamma=\alpha$) and polar ($\beta$) mixtures are included  with time-reversal (TR) compatible phases and small real coefficients $\delta^\gamma_i$'s. Each $\delta$ is distinct for different states except in the same IR (e.g., $T_\pm$ of [001] in Table~\ref{tab:IR}).  Without crystal anisotropy, its difference between $S$ and $T$ is due to the exchange part (as opposed to the direct Coulomb), and  $\delta^\beta_{0}=0$ for $T$ states due to Pauli exclusion.  $\delta^\gamma_i$'s can be obtained perturbatively due to the $H_{\rm int-d}$ interaction (see Appendix~\ref{app:bbb}). The magnitude of a general $\gamma_X$ mixture into the unperturbed $\alpha_{A_1}$ state is on the order of exchange interaction between two-electron states, $\alpha_{A_1}(\mathbf{r}_1) \beta_{A_1}(\mathbf{r}_2)$ and $\alpha_{A_1}(\mathbf{r}_2) \gamma_{X}(\mathbf{r}_1)$, divided by the energy difference of these two states.

The perturbation theory for relaxation matrix elements throughout Eqs.~(\ref{eq:S-T0-001})-(\ref{eq:T-barT-110}) then proceeds in a straightforward manner following the  above prescriptions.
We find the symmetries of the relevant basis states and interaction operators, and determine various perturbation integrals, as shown in the technical details  in Appendix~\ref{app:ccc}. Here we illustrate the key common aspects by analyzing the representative Eq.~(\ref{eq:S-T0-001}) in more detail.
Both terms require inter-donor interaction amounting to two overlap factors, so neither of them may be neglected  a priori. As shown in Eq.~(\ref{eq:SOCz_needed}), SOC is necessary. We find that it is dominated by donor SOC ($\sim\lambda_{\rm soc} \mathbf{L}\cdot \mathbf{s}$, where  $\mathbf{L}$ denotes the operator conjugating to the spin in SOC, i.e., $\mathbf{L}\propto \bm\nabla V\times \mathbf{p}$).  The donor $L_z$ and $\epsilon_{xx}-\epsilon_{yy}$ bring $\Delta_1$ band back to itself allowing the remaining inter-donor coupling (we note again that  the second electron of $\mathbf{r}_2$ in the $H_{\rm int-d}$ integral is always in the donor ground states).  $\epsilon_{xx}-\epsilon_{yy}$, moreover, connects two available valley configurations comprising $\Delta_1$-$1s$ states.  Together, the $\gamma=\alpha$ term in Eq.~(\ref{eq:S-T0-001}) contains a perturbation expansion,
\begin{widetext}
\begin{eqnarray}
\langle T_0| \epsilon_{xx}-\epsilon_{yy} | S \rangle_{[001]}^{(1)}
\!\propto\!\! \sum_{\nu=1s,3d_{\pm1}}\!\!\!\frac{\langle \alpha^{T_0}_{\Delta_1\!,1s\!,A_1}| \overline{H}_{\rm int\!-\!d}|\alpha^{T_0}_{\Delta_1\!,\nu\!,T_{2z}}\rangle \langle \alpha_{\Delta_1\!,\nu\!,T_{2z}}| L_z|\alpha_{\Delta_1\!,1s\!,E^I_z}\rangle \langle \alpha_{\Delta_1\!,1s\!,E^I_z}| \epsilon_{xx}\!-\!\epsilon_{yy} | \alpha_{\Delta_1\!,1s\!,A_1}\rangle}
{(\mathcal{E}_{\Delta_1,1s,A_1}- \mathcal{E}_{\Delta_1,\nu,T_2}) (\mathcal{E}_{\Delta_1,1s,A_1}- \mathcal{E}_{\Delta_1,1s,E})}
,
\label{eq:S-T0-001-1}
\end{eqnarray}
\end{widetext}
plus another one with reversed ordering of interactions, where $\overline{H}_{\rm int-d}$ couples $A_1$ and $T_{2z}$ envelopes in the same donor [as expected from Eq.~(\ref{eq:001_Td_mix-alphaUp})] of the $S$ instead of $T_0$ state. That leaves only the exchange part of $H_{\rm int-d}$ effective ($\sim J_{A_1T_2}$). $\mathcal{E}_{\Delta_1,1s,A_1}- \mathcal{E}_{\Delta_1,1s,E/T_2}(\mathcal{E}_{\Delta_1,3d_{\pm1},T_2})\approx -12 (40)$ meV \cite{Aggarwal_PR65}, and $\langle E^I_z| \epsilon_{xx}\!-\!\epsilon_{yy} |A_1\rangle=\sqrt{2/3} \Xi_u$ with $\Xi_u\approx8.77$ eV \cite{Yu_Cardona_Book}. Two major donor SOC couplings emerge: one is between two $1s$ configurations and relates to the donor spin splitting $\Delta^{\rm dnr}_{\rm SOC}$($\sim$ 0.03, 0.1, 0.3 meV for P, As, Sb donors respectively;  this impurity core effect is deduced from experiments and goes beyond effective mass approximation) \cite{Castner_PR67,Song_PRL14}. The other is between $1s$ and $3d_{\pm1}$ where $\langle 3d_{\pm 1}|L_z|1s\rangle\neq 0$ within a \textit{single} ($x$ or $y$) valley due to anisotropy of the $1s$ envelope \cite{Kohn_PR55}.
We find the $\nu=1s$ component can be safely used for an order of magnitude estimate \cite{footnote_3d}. For the $\gamma=\beta$ term in Eq.~(\ref{eq:S-T0-001}), the $J_{A_1T_2}$ factor is replaced by an exchange term with the polar state $\beta_{A_1}\beta_{T_2}$ multiplied by another overlap factor which, therefore, has the same order of magnitude. By re-ordering the interactions, we see that two other perturbation expansions of  similar or smaller magnitudes are allowed [shown in Eqs.~(\ref{eq:S-T0-001-2}) and (\ref{eq:S-T0-001-3})].

More perturbation terms of similar magnitudes exist, representing the combined interaction operators. The typical example is the \textit{Yafet} term in the E-Y spin flip mechanism (purely opposite spin states coupled by the SOC part of e-ph interaction) \cite{Yafet_SSP63}. Others include phonon-modulated exchange interaction. It is not practical to enumerate each term, however. Even in the much simpler pure bulk Si,  numerous comparable leading-order terms contribute \cite{Song_PRB12}. In the single-donor relaxation, for example, the Yafet term is not considered \cite{Hasegawa_PR60_Roth_PR60_Roth_LincolnLab60}. More importantly, it is also not essential to do so since these terms possess the same symmetry dependence on operator components (e.g., $\epsilon_{ij}, s_i$) according to the method of invariants \cite{Luttinger_PR56, Bir_Pikus_Book}, justifying this treatment. The net numerical prefactors are difficult to evaluate exactly, and should be left to be extracted  experimentally aided by our transparent expressions.

\section{Quantitative analysis and discussion}\label{sec:quantitative}

Following the derivation shown above, we obtain the leading-order magnitude $|M|$ for relaxation channels $\langle T_0|...|S\rangle_{[110]}$, $\langle T_\pm|...|S\rangle_{[111]n}$, and $\langle T_+ +T_-|...|S\rangle_{[110]}$ (indexed by $\kappa=2,3,4$ whereas $\langle T_0|...|S\rangle_{[001]}$ by $\kappa=1$) via donor SOC,
\begin{eqnarray}\label{eq:M}
M_{\kappa,\lambda} &\approx& \epsilon^{ph}_{\kappa,\lambda} \mathcal{F}_\kappa\frac{J_{A_1 T_2}  \sqrt{\frac{2}{3}} \Xi_u  \Delta^{\rm dnr}_{\rm SOC}  }{\Delta \mathcal{E}_1 \Delta \mathcal{E}_2},
\end{eqnarray}
where $\lambda$ is the phonon mode, $\Delta \mathcal{E}_{1(2)}=-12$ meV, channel-dependent factor $\mathcal{F}_{1,2,3,4}= 1, \frac{1}{\sqrt{2}}, \frac{3\sqrt{2}i}{8}, \frac{2}{3\sqrt{2}}$, respectively,   and
$\epsilon^{ph}_{\kappa,\lambda} = \tilde{\epsilon}_{\kappa,\lambda} \sqrt{\hbar (n_{q_\lambda}+1) /2 \rho \omega_{\lambda} (q)} $.
$ \tilde{\epsilon}_{\kappa,\lambda}$ for $\kappa= 1,2$ is $\frac{1}{2}(\tilde{\epsilon}_{xx}-\tilde{\epsilon}_{yy})$, for $\kappa= 3$ is $\frac{1}{3\sqrt{2}} (e^{i\frac{\pi}{6}}\tilde{\epsilon}_{xx}-i\tilde{\epsilon}_{yy} - e^{-i\frac{\pi}{6}}\tilde{\epsilon}_{zz} +e^{-i\frac{\pi}{6}}\tilde{\epsilon}_{xy}+i\tilde{\epsilon}_{xz} -e^{i\frac{\pi}{6}}\tilde{\epsilon}_{yz} )$ and for $\kappa= 4$ is $\tilde{\epsilon}_{zz} +\frac{1}{2}(\tilde{\epsilon}_{xx}+\tilde{\epsilon}_{yy})$ where $\tilde{\epsilon}_{ij}\equiv i(q_i\xi_j+q_j\xi_i)/2$, and then phonon polarization $\bm{\xi}(\mathbf{q})$ is projected   into $\lambda=$ LA, TA$_1$, TA$_2$  by elastic continuum approximation \cite{Ehrenreich_PR56}. $\rho=2.33$ g/cm$^3$, $\hbar\omega_\lambda=\hbar v_\lambda q$ equals the relaxation energy, $v_{\rm LA(TA)}=8.7(5)\times 10^5$ cm/s, and $n_q$ is the phonon distribution.
Equation~(\ref{eq:M}) is not expected to be exact as discussed above. Our goal is to provide the leading-order estimate for $M$ and its functional dependence on $J$ and temperature. We find for $\langle T_\pm|...|S\rangle_{[001]}$ and $\langle T_+-T_-|...|S\rangle_{[110]}$ the leading-order term vanishes by symmetry and effective-mass analysis, as shown in Appendix~\ref{app:ccc}, and our mechanism is overshadowed by hyperfine-induced relaxation \cite{Borhani_PRB10}. $M$ between $T$ states are also obtained with a critical difference regarding the inter-donor coupling. Unlike between $S$ and $T$ states, its effect stems from the anisotropic SOC part of $H_{\rm int-d}$ and scales down roughly by a large factor $|J/I|$ where $I=\mathcal{E}_{T_0}-\mathcal{E}_{T_\pm}$.

Our result differs from the existing hyperfine-coupling theory on two-donor relaxation in Refs.~\cite{Sugihara_JPCS68} and \cite{Borhani_PRB10} in a critical way that the spin mixing scales with $|$SOC coupling/spinless coupling$|$ ($\sim$ 0.03 meV/12 meV) instead of $|$hyperfine($A$)/exchange($J$)$|$ ($A\sim 0.2 \mu$eV and $J$ is on the order of 1 meV for $d\sim 8$ nm). Thus our mechanism provides a comparable or stronger relaxation channel at the higher end of the two-qubit coupling regime ($\le 10$ nm) \cite{Kane_Nat98, Koiller_PRL02}, and is therefore extremely relevant for the quantum computing platforms involving $S$-$T$ donor-pair qubits \cite{Klymenko_JPCM14, Dehollain_PRL14, Gonzales-Zalba_NanoLett14}.  Also our stronger scaling on $J$ (by $J^2$) offers a clear experimental distinction between the two mechanisms.

\begin{figure}[!htbp]
\includegraphics[width=7cm]{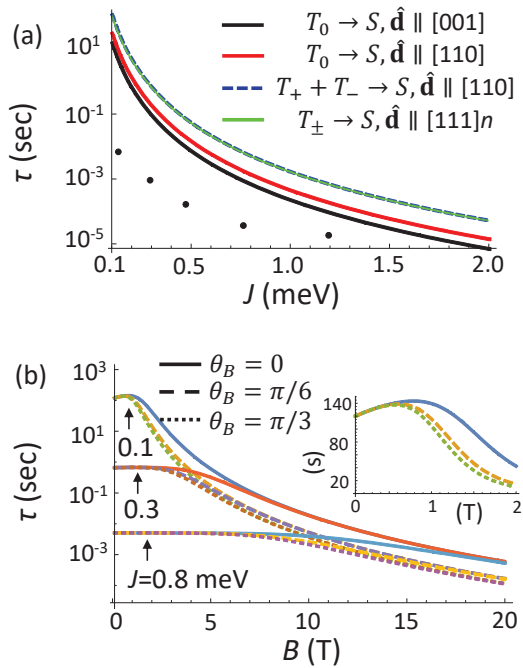}
\caption { (a) Relaxation time $\tau$ for four leading-order channels driven by P donor SOC as 0.1 meV$<J<2$ meV. The numerical result for $T_0\rightarrow S$, $\hat{\mathbf{d}}\|[001]$ (from \cite{Borhani_PRB10}) is marked by solid dots. (b) $\tau$ of $T_+\rightarrow S$ and $\hat{\mathbf{d}}\| [111]n$ for three different $J$'s as  $B<20$ T. The inset shows an increasing $\tau$ at $g\mu_B B\lesssim k_B T$, a feature of finite exchange coupling. $T=1$K and the single-electron relaxation uses experimental parameters from  \cite{Wilson_PR61}.} \label{fig}
\end{figure}

Figure~\ref{fig}(a) quantifies the relaxation time $\tau$ for exchange coupling between 0.1 and 2 meV, by $\tau^{-1}= \frac{2\pi}{\hbar} \int \frac{d^3 q}{(2\pi)^3} \sum_{ \lambda}|M_{\lambda}|^2 \delta(J-\hbar v_\lambda q_\lambda)$ via Eq.~(\ref{eq:M}) and $\int d\Omega_{\mathbf{q}} |\tilde{\epsilon}_{\kappa,\lambda}|^2$ \cite{footnote_tau}. $\tau\propto J^{-5}$ ($k_B T\ll J$) or $T J^4$ ($k_B T\gg J$).  The dependence on the phonon mode, which is rigorous from Eqs.~(\ref{eq:S-T0-001})-(\ref{eq:T-barT-110}), is made explicit here. The numerical factor due to inter-donor interaction, however, is crudely averaged as exchange splitting $J$. Various exchange terms from different perturbation expansions [as discussed between Eqs.~(\ref{eq:S-T0-001-1}) and (\ref{eq:M})] lead to superposed oscillations over donor vector $\mathbf{d}$,  unrealistic for a perturbation calculation like ours to specify.
To have a rough idea about the relative trends between our mechanism and the hyperfine-induced one, we also add the hyperfine result from Ref.~\cite{Borhani_PRB10} in Fig.~\ref{fig}(a). Although the spin mixing is generally stronger in our mechanism, the phonon-mode related factors $ \tilde{\epsilon}_{\kappa,\lambda}$  reduce our relaxation rates somewhat relatively. The ratio of relaxation time over exchange gate time goes as $J \tau/\pi\sim 1/J^4$ and decreases sharply with $J$. However, even the smallest ratio in this figure, for $T_0\rightarrow S, \hat{\mathbf{d}}\|[001]$ at $J=2$ meV, is about 7 million, well satisfying the need for quantum error correction.  We remark that for As and Sb donors, the crossover between SOC and hyperfine mechanisms occurs at smaller $J$.  In this regime, our mechanism indicates that quantum computation using As \cite{Gonzales-Zalba_NanoLett14, Lo_APL14} or Sb \cite{Bradbury_06PRL} donors has a much shorter intrinsic relaxation time compared to that of P donors. Note that while some uncertainty in the overall prefactors may raise or lower our curves in Fig.~\ref{fig}(a),  the parametrically dominant $J^5$ dependence is robust and independent of quantitative details, clearly establishing the dominance of our mechanism for stronger inter-donor exchange coupling.


Finally,  TR  forbids spin relaxation by SOC on a single donor strictly at zero  magnetic ($B$) field \cite{Hasegawa_PR60_Roth_PR60_Roth_LincolnLab60} whereas it does not affect any transition among the lowest two-electron states. As a result, the application of $B$ field has the most effect on driving one-electron spin flips, which dominate electron relaxation when $g\mu_B B\gtrsim J$. Indeed, we see a crossover behavior with increasing $B$ in Fig.~\ref{fig}(b) for an $S\rightarrow T_-$ relaxation. The crossover is much sooner for $T_0 \rightarrow T_-$ as $g\mu_B B\sim I$. In fact, since the $B$ power-law dependence for single donors and our $J$ or $I$ dependence are exactly the same for any temperature, the crossover occurs at a fixed $g\mu_B B/J(I)$ ratio depending only on the donor type. This is an important finding independent of quantitative details.

\section{Summary}\label{sec:summary}

In conclusion, we have established that two-electron relaxation in Si donor pairs driven by the intrinsic SOC is important, despite  the decade-long perception that the SOC in this system is too weak and only hyperfine-induced relaxation needs to be taken into account \cite{Sugihara_JPCS68, Borhani_PRB10}. It is crucial that we identify that the SOC from the central cell region of the donor core, not from the ion Coulomb potential or bulk Si, is the dominant source. The central cell SOC stems from the fast-varying core potential (which the smooth Coulomb potential  does not exhibit) and at the same time breaks the Elliott-Yafet cancellation that comes with the bulk spin-phonon interaction in Si. We unravel the intricate relaxation matrix elements which invoke inter-donor, SOC and e-ph interaction on top of the Si crystal and ionized donor potentials, by  utilizing the exact symmetries possessed by the two-donor system Hamiltonian and also the approximate ones by the individual donors and bulk crystal (resulting in  various $\Delta$ bands at six points of the $\Delta$ stars). In this way, we are able to finally obtain a relatively simple and robust dependence of the relaxation rates on the donor alignment and exchange interactions, phonon modes and deformation potentials, and SOC constants, which cannot be ascertained by brute-force numerical simulations. Our mechanism dominates the qubit relaxation when the two-donor separation is relatively small or the inter-qubit coupling is large, and is especially important for donor-based singlet-triplet qubits. Therefore, this study is essential for any  serious quantum computing  proposals based on Si-donor spin qubits.

Together with the hyperfine-driven two-electron relaxation in the coupled donor systems that has been the only known mechanism for half a century \cite{Sugihara_JPCS68, Borhani_PRB10},  we now have a full theory for the noise-free intrinsic relaxation limit of the two-donor spin states.  When both hypefine and SOC are included, the effect should be of higher order and thus comparably negligible (i.e. our effect and hyperfine effect are additive to leading order). Future developments incorporating charge noise, realistic interface or quantum well confinement potentials can expand on our results and provide quantitative estimates for  comparisons with experimental data when they become available in the strong donor-coupling regime. On the experimental side, the SOC and hyperfine related relaxation mechanisms can be disentangled and compared by varying donor separation and alignment direction, and by going over different donor species (P, As, Sb).

\section*{ACKNOWLEDGMENTS}

This work is supported by Laboratory for Physical Sciences and Microsoft.

\appendix

\section{symmetry identifications of the inter-donor coupled mixture}\label{app:aaa}

\begin{table} [!htbp]
\caption{\label{tab:symmetry_matrices_001}
Transformation matrices of spatial and TR symmetry operations for a two-donor system aligned   along the [001] direction, i.e., $D_{2d}$ group, upon $T_d$ group IRs and spin basis states. For concreteness, we specify for the  degenerate IRs that, $E^I_z\sim x^2-y^2$, $E^{I\!I}_z\sim (x^2+y^2-2z^2)/\sqrt{3}$; $T_{1x}\sim R_x, T_{1y}\sim R_y, T_{1z}\sim R_z$; and $T_{2x}\sim x, T_{2y}\sim y, T_{2z}\sim z$  in terms of their transformation properties.
}
\renewcommand{\arraystretch}{2.5}
\tabcolsep=0.02 cm
\begin{tabular}{ccccccc}
 \hline\hline
            &   $A_1$   &  $A_2$ & $E^I, E^{I\!I}$ & $T^x_1, T^y_1, T^z_1$ & $T^x_2, T^y_2, T^z_2$ & $\uparrow_z,\downarrow_z$\\
               \hline
$C_{2z}$    &   1       &  1     & $\! \left( \!\!\renewcommand{\arraycolsep}{2.pt}\renewcommand{\arraystretch}{0.8} \begin{array}{cc}1&0\\0&1\end{array} \!\right)$  & $\!\!\! \left( \!\! \renewcommand{\arraycolsep}{2.pt}\renewcommand{\arraystretch}{0.8} \begin{array}{ccc}-1&0&0 \\0&-1&0 \\ 0&0&1 \end{array} \!\!\right)$  & $\!\!\! \left( \!\!\renewcommand{\arraycolsep}{2.pt}\renewcommand{\arraystretch}{0.8}  \begin{array}{ccc}-1&0&0 \\0&-1&0 \\ 0&0&1 \end{array} \!\!\right)$  & $\!\!\! \left( \!\renewcommand{\arraycolsep}{2.pt}\renewcommand{\arraystretch}{0.8}  \begin{array}{cc}-i&0\\0&i\end{array} \!\right)$
\\
$\rho_{[1-10]}$ & 1 & $-1$ & $\!\!\! \left( \!\!\renewcommand{\arraycolsep}{2.pt}\renewcommand{\arraystretch}{0.8}  \begin{array}{cc}-1&0\\0&1\end{array} \!\!\right)$  &
$\!\!\! \left( \!\!\renewcommand{\arraycolsep}{2.pt}\renewcommand{\arraystretch}{0.8}  \begin{array}{ccc}0&-1&0 \\-1&0&0 \\ 0&0&-1 \end{array} \!\!\right)$  &
$\!\!\! \left( \!\!\renewcommand{\arraycolsep}{2.pt}\renewcommand{\arraystretch}{0.8}  \begin{array}{ccc}0&1&0 \\1&0&0 \\ 0&0&1 \end{array} \!\!\right)$  &
 $\!\!\! \left( \!\renewcommand{\arraycolsep}{2.pt}\renewcommand{\arraystretch}{0.8} \begin{array}{cc}0&\frac{1-i}{\sqrt{2}}\\-\frac{1+i}{\sqrt{2}}&0\end{array} \!\right)$
\\
$\rho_{[110]}$ & 1 & $-1$ & $\!\!\! \left( \!\!\renewcommand{\arraycolsep}{2.pt}\renewcommand{\arraystretch}{0.8}  \begin{array}{cc}-1&0\\0&1\end{array} \!\right)$  & $\!\!\! \left( \!\!\renewcommand{\arraycolsep}{2.pt}\renewcommand{\arraystretch}{0.8}  \begin{array}{ccc}0&1&0 \\1&0&0 \\ 0&0&-1 \end{array} \!\!\right)$  & $\!\!\! \left( \!\!\renewcommand{\arraycolsep}{2.pt}\renewcommand{\arraystretch}{0.8} \begin{array}{ccc}0&-1&0 \\-1&0&0 \\ 0&0&1 \end{array} \!\!\right)$  & $\!\!\! \left( \!\renewcommand{\arraycolsep}{2.pt}\renewcommand{\arraystretch}{0.8} \begin{array}{cc}0&-\frac{1+i}{\sqrt{2}}\\ \frac{1-i}{\sqrt{2}}&0\end{array} \!\right)$
\\
$C_{2x}$    &   1       &  1     & $\!\!\! \left( \!\!\renewcommand{\arraycolsep}{2.pt}\renewcommand{\arraystretch}{0.8} \begin{array}{cc}1&0\\0&1\end{array} \!\!\right)$  &
$\!\!\! \left( \!\!\renewcommand{\arraycolsep}{2.pt}\renewcommand{\arraystretch}{0.8} \begin{array}{ccc}1&0&0 \\0&-1&0 \\ 0&0&-1 \end{array} \!\!\right)$  &
$\!\!\! \left( \!\!\renewcommand{\arraycolsep}{2.pt}\renewcommand{\arraystretch}{0.8}\begin{array}{ccc}1&0&0 \\0&-1&0 \\ 0&0&-1 \end{array} \!\!\right)$  &
$\!\!\! \left( \!\renewcommand{\arraycolsep}{2.pt}\renewcommand{\arraystretch}{0.8} \begin{array}{cc}0&-i\\-i&0\end{array} \!\right)$
\\
$C_{2y}$    &   1       &  1     & $\!\!\! \left( \!\!\renewcommand{\arraycolsep}{2.pt}\renewcommand{\arraystretch}{0.8} \begin{array}{cc}1&0\\0&1\end{array} \!\!\right)$  & $\!\!\! \left( \!\!\renewcommand{\arraycolsep}{2.pt}\renewcommand{\arraystretch}{0.8}\begin{array}{ccc}-1&0&0 \\0&1&0 \\ 0&0&-1 \end{array} \!\!\right)$  & $\!\!\! \left( \!\!\renewcommand{\arraycolsep}{2.pt}\renewcommand{\arraystretch}{0.8} \begin{array}{ccc}-1&0&0 \\0&1&0 \\ 0&0&-1 \end{array} \!\!\right)$  & $\!\!\! \left( \!\!\renewcommand{\arraycolsep}{2.pt}\renewcommand{\arraystretch}{0.8} \begin{array}{cc}0&-1\\1&0\end{array} \!\!\right)$
\\
$S^+_{4z}$ & 1 &$-1$ & $\!\!\! \left( \!\!\renewcommand{\arraycolsep}{2.pt}\renewcommand{\arraystretch}{0.8} \begin{array}{cc}-1&0\\0&1\end{array} \!\!\right)$  & $\!\!\! \left( \!\!\renewcommand{\arraycolsep}{2.pt}\renewcommand{\arraystretch}{0.8} \begin{array}{ccc}0&1&0 \\-1&0&0 \\ 0&0&1 \end{array} \!\!\right)$  & $\!\!\! \left( \!\!\renewcommand{\arraycolsep}{2.pt}\renewcommand{\arraystretch}{0.8} \begin{array}{ccc}0&-1&0 \\1&0&0 \\ 0&0&-1 \end{array} \!\!\right)$  & $\!\!\! \left( \!\!\renewcommand{\arraycolsep}{2.pt}\renewcommand{\arraystretch}{0.8} \begin{array}{cc}-\frac{1+i}{\sqrt{2}}&0\\ 0&\frac{i-1}{\sqrt{2}}\end{array} \!\!\right)$
\\
$S^-_{4z}$ & 1 & $-1$ & $\!\!\! \left( \!\!\renewcommand{\arraycolsep}{2.pt}\renewcommand{\arraystretch}{0.8}\begin{array}{cc}-1&0\\0&1\end{array} \!\!\right)$  & $\!\!\! \left( \!\!\renewcommand{\arraycolsep}{2.pt}\renewcommand{\arraystretch}{0.8} \begin{array}{ccc}0&-1&0 \\1&0&0 \\ 0&0&1 \end{array} \!\!\right)$  & $\!\!\! \left( \!\!\renewcommand{\arraycolsep}{2.pt}\renewcommand{\arraystretch}{0.8}\begin{array}{ccc}0&1&0 \\-1&0&0 \\ 0&0&-1 \end{array} \!\!\right)$  & $\!\!\! \left( \!\!\renewcommand{\arraycolsep}{2.pt}\renewcommand{\arraystretch}{0.8} \begin{array}{cc}\frac{1-i}{\sqrt{2}}&0\\ 0&\frac{1+i}{\sqrt{2}}\end{array} \!\!\right)$
\\
TR    &   1       &  1     & $\!\!\! \left( \!\!\renewcommand{\arraycolsep}{2.pt}\renewcommand{\arraystretch}{0.8}\begin{array}{cc}1&0\\0&1\end{array} \!\!\right)$  & $\!\!\! \left( \!\!\renewcommand{\arraycolsep}{2.pt}\renewcommand{\arraystretch}{0.8}\begin{array}{ccc}-1&0&0 \\0&-1&0 \\ 0&0&-1 \end{array} \!\!\right)$  & $\!\!\! \left( \!\!\renewcommand{\arraycolsep}{2.pt}\renewcommand{\arraystretch}{0.8}\begin{array}{ccc}-1&0&0 \\0&-1&0 \\ 0&0&-1 \end{array} \!\!\right)$  & $\!\!\! \left( \!\!\renewcommand{\arraycolsep}{2.pt}\renewcommand{\arraystretch}{0.8}\begin{array}{cc}0&-i\\i&0\end{array} \!\!\right)$
\\
\hline\hline
\end{tabular}
\end{table}

In this appendix, we present the allowed IRs of mixture for each of the donor-alignment  directions $\hat{\mathbf{d}}$ to be [001], [111], and [110]. The leading-order mixtures of interest are such that one of the two electron states is the same as that in the unperturbed Heitler-London ground states.  The mixed components and their coefficients have to obey all the spatial and time reversal (TR) symmetry of the two-donor systems.

\begin{table} [!htbp]
\caption{\label{tab:symmetry_matrices_111}
Transformation matrices of spatial and TR symmetry operations for a two-donor system aligned  along the [111] direction, i.e., $C_{3v}$ group without inversion or $D_{3d}$ group with inversion (which connects $\alpha$ and $\beta$ donors and keeps the spin direction),  upon $T_d$ group IRs and spin basis states.
}
\renewcommand{\arraystretch}{2.5}
\tabcolsep=0.02 cm
\begin{tabular}{ccccccc}
 \hline\hline
            &   $A_1$   &  $A_2$ & $E^I, E^{I\!I}$ & $T^x_1, T^y_1, T^z_1$ & $T^x_2, T^y_2, T^z_2$ & $\uparrow_{[111]},\downarrow_{[111]}$\\
               \hline
$C^+_{3[111]}$    &   1       &  1     & $\!\!\! \left( \!\!\renewcommand{\arraycolsep}{2.pt}\renewcommand{\arraystretch}{0.8} \begin{array}{cc}-\frac{1}{2}&-\frac{\sqrt{3}}{2}\\\frac{\sqrt{3}}{2}&-\frac{1}{2}\end{array} \!\!\right)$  & $\!\!\! \left( \!\!\renewcommand{\arraycolsep}{2.pt}\renewcommand{\arraystretch}{0.8}  \begin{array}{ccc}0&0&1 \\1&0&0 \\ 0&1&0 \end{array} \!\!\right)$  & $\!\!\! \left( \!\!\renewcommand{\arraycolsep}{2.pt}\renewcommand{\arraystretch}{0.8}  \begin{array}{ccc}0&0&1 \\1&0&0 \\ 0&1&0 \end{array} \!\!\!\right)$  & $\!\!\! \!\!\left( \!\!\renewcommand{\arraycolsep}{1.pt}\renewcommand{\arraystretch}{0.8}  \begin{array}{cc} e^{\frac{-i\pi}{3}}&0\\0&e^{\frac{i\pi}{3}}\end{array} \!\!\right)$
\\
$C^-_{3[111]}$    &   1       &  1     & $\!\!\! \left( \!\!\renewcommand{\arraycolsep}{2.pt}\renewcommand{\arraystretch}{0.8} \begin{array}{cc}-\frac{1}{2}&\frac{\sqrt{3}}{2}\\-\frac{\sqrt{3}}{2}&-\frac{1}{2}\end{array} \!\!\right)$  & $\!\!\! \left( \!\!\renewcommand{\arraycolsep}{2.pt}\renewcommand{\arraystretch}{0.8}  \begin{array}{ccc}0&1&0 \\0&0&1 \\ 1&0&0 \end{array} \!\!\right)$  & $\!\!\! \left( \!\!\renewcommand{\arraycolsep}{2.pt}\renewcommand{\arraystretch}{0.8}  \begin{array}{ccc}0&1&0 \\0&0&1 \\ 1&0&0 \end{array} \!\!\right)$  & $\!\!\! \!\!\left( \!\!\renewcommand{\arraycolsep}{1.pt}\renewcommand{\arraystretch}{0.8}  \begin{array}{cc}e^{\frac{i\pi}{3}}&0\\0&e^{\frac{-i\pi}{3}}\end{array} \!\!\right)$
\\
$\rho_{[1-10]}$ & 1 & $-1$ & $\!\!\! \left( \!\!\renewcommand{\arraycolsep}{2.pt}\renewcommand{\arraystretch}{0.8}  \begin{array}{cc}-1&0\\0&1\end{array} \!\!\right)$  & $\!\!\! \left( \!\!\renewcommand{\arraycolsep}{1.pt}\renewcommand{\arraystretch}{0.8} \begin{array}{ccc}0&-1&0 \\-1&0&0 \\ 0&0&-1 \end{array} \!\!\right)$  & $\!\!\! \left( \!\!\renewcommand{\arraycolsep}{2.pt}\renewcommand{\arraystretch}{0.8} \begin{array}{ccc}0&1&0 \\1&0&0 \\ 0&0&1 \end{array} \!\!\right)$  & $\!\!\!\!\! \left( \!\!\renewcommand{\arraycolsep}{1.pt}\renewcommand{\arraystretch}{0.8} \begin{array}{cc}0&\frac{1-i}{\sqrt{2}}\\ -\frac{1+i}{\sqrt{2}}&0\end{array} \!\!\right)$
\\
$\rho_{[10-1]}$ & 1 & $-1$ & $\! \left( \!\!\renewcommand{\arraycolsep}{1.pt}\renewcommand{\arraystretch}{0.8} \begin{array}{cc}\frac{1}{2}&-\frac{\sqrt{3}}{2}\\-\frac{\sqrt{3}}{2}&-\frac{1}{2}\end{array} \!\!\right)$  & $\!\!\! \left( \!\!\renewcommand{\arraycolsep}{1.pt}\renewcommand{\arraystretch}{0.8}  \begin{array}{ccc}0&0&-1 \\0&-1&0 \\ -1&0&0 \end{array} \!\!\right)$  & $\!\!\! \left( \!\!\renewcommand{\arraycolsep}{2.pt}\renewcommand{\arraystretch}{0.8}  \begin{array}{ccc}0&0&1 \\0&1&0 \\ 1&0&0 \end{array} \!\!\right)$  & $\!\!\!\!\! \left( \!\!\renewcommand{\arraycolsep}{2.pt}\renewcommand{\arraystretch}{0.8} \begin{array}{cc}0&-i\\ -i&0\end{array} \!\!\right)$
\\
$\rho_{[01-1]}$ & 1 & $-1$ & $\!\!\! \left( \!\!\renewcommand{\arraycolsep}{2.pt}\renewcommand{\arraystretch}{0.8} \begin{array}{cc}\frac{1}{2}&\frac{\sqrt{3}}{2}\\\frac{\sqrt{3}}{2}&-\frac{1}{2}\end{array} \!\!\right)$  & $\!\!\! \left( \!\!\renewcommand{\arraycolsep}{1.pt}\renewcommand{\arraystretch}{0.8}  \begin{array}{ccc}-1&0&0 \\0&0&-1 \\ 0&-1&0 \end{array} \!\!\right)$  & $\!\!\! \left( \!\!\renewcommand{\arraycolsep}{2.pt}\renewcommand{\arraystretch}{0.8} \begin{array}{ccc}1&0&0 \\0&0&1 \\ 0&1&0 \end{array} \!\!\right)$  & $\!\!\!\!\! \left( \!\!\renewcommand{\arraycolsep}{1.pt}\renewcommand{\arraystretch}{0.8}  \begin{array}{cc}0&-\frac{1+i}{\sqrt{2}}\\ \frac{1-i}{\sqrt{2}}&0\end{array} \!\!\right)$
\\
$i$ & 1 & $-1$ & $\!\!\! \left( \!\!\renewcommand{\arraycolsep}{2.pt}\renewcommand{\arraystretch}{0.8} \begin{array}{cc}\frac{1}{2}&\frac{\sqrt{3}}{2}\\\frac{\sqrt{3}}{2}&-\frac{1}{2}\end{array} \!\!\right)$  & $\!\!\! \left( \!\!\renewcommand{\arraycolsep}{1.pt}\renewcommand{\arraystretch}{0.8}  \begin{array}{ccc}-1&0&0 \\0&0&-1 \\ 0&-1&0 \end{array} \!\!\right)$  & $\!\!\! \left( \!\!\renewcommand{\arraycolsep}{2.pt}\renewcommand{\arraystretch}{0.8} \begin{array}{ccc}1&0&0 \\0&0&1 \\ 0&1&0 \end{array} \!\!\right)$  & $\!\!\!\!\! \left( \!\!\renewcommand{\arraycolsep}{1.pt}\renewcommand{\arraystretch}{0.8} \begin{array}{cc}0&-\frac{1+i}{\sqrt{2}}\\ \frac{1-i}{\sqrt{2}}&0\end{array} \!\!\right)$
\\
TR    &   1       &  1     & $\!\!\! \left( \!\!\renewcommand{\arraycolsep}{2.pt}\renewcommand{\arraystretch}{0.8}  \begin{array}{cc}1&0\\0&1\end{array} \!\!\right)$  & $\!\!\! \left( \!\!\renewcommand{\arraycolsep}{1.pt}\renewcommand{\arraystretch}{0.8} \begin{array}{ccc}-1&0&0 \\0&-1&0 \\ 0&0&-1 \end{array} \!\!\right)$  & $\!\!\! \left( \!\!\renewcommand{\arraycolsep}{1.pt}\renewcommand{\arraystretch}{0.8} \begin{array}{ccc}-1&0&0 \\0&-1&0 \\ 0&0&-1 \end{array} \!\!\right)$  & $\!\!\! \left( \!\!\renewcommand{\arraycolsep}{2.pt}\renewcommand{\arraystretch}{0.8} \begin{array}{cc}0&-i\\i&0\end{array} \!\!\right)$
\\
\hline\hline
\end{tabular}
\end{table}

\begin{table} [!htbp]
\caption{\label{tab:symmetry_matrices_110}
Transformation matrices of spatial and TR symmetry operations for a two-donor system aligned  along the [110] direction, i.e., $C_{2v}$ group, upon $T_d$ group IRs and spin basis states.
}
\renewcommand{\arraystretch}{2.0}
\tabcolsep=0.04 cm
\begin{tabular}{ccccccc}
 \hline\hline
            &   $A_1$   &  $A_2$ & $E^I, E^{I\!I}$ & $T^x_1, T^y_1, T^z_1$ & $T^x_2, T^y_2, T^z_2$ & $\uparrow_{[110]},\downarrow_{[110]}$\\
               \hline
$C_{2z}$    &   1       &  1     & $\!\!\! \left( \!\!\renewcommand{\arraycolsep}{2.pt}\renewcommand{\arraystretch}{0.8}\begin{array}{cc}1&0\\0&1\end{array} \!\!\right)$  & $\!\!\! \left( \!\!\renewcommand{\arraycolsep}{2.pt}\renewcommand{\arraystretch}{0.8} \begin{array}{ccc}-1&0&0 \\0&-1&0 \\ 0&0&1 \end{array} \!\!\right)$  & $\!\!\! \left( \!\!\renewcommand{\arraycolsep}{2.pt}\renewcommand{\arraystretch}{0.8}\begin{array}{ccc}-1&0&0 \\0&-1&0 \\ 0&0&1 \end{array} \!\!\right)$  & $\!\!\! \left( \!\!\renewcommand{\arraycolsep}{2.pt}\renewcommand{\arraystretch}{0.8} \begin{array}{cc}0&-i\\-i&0\end{array} \!\!\right)$
\\
$\rho_{[110]}$ & 1 & -1 & $\!\!\! \left( \!\!\renewcommand{\arraycolsep}{2.pt}\renewcommand{\arraystretch}{0.8} \begin{array}{cc}-1&0\\0&1\end{array} \!\!\right)$  & $\!\!\! \left( \!\!\renewcommand{\arraycolsep}{2.pt}\renewcommand{\arraystretch}{0.8}\begin{array}{ccc}0&1&0 \\1&0&0 \\ 0&0&-1 \end{array} \!\!\right)$  & $\!\!\! \left( \!\!\renewcommand{\arraycolsep}{2.pt}\renewcommand{\arraystretch}{0.8}\begin{array}{ccc}0&-1&0 \\-1&0&0 \\ 0&0&1 \end{array} \!\!\right)$  & $\!\!\! \left( \!\!\renewcommand{\arraycolsep}{2.pt}\renewcommand{\arraystretch}{0.8} \begin{array}{cc}-i&0\\ 0&i\end{array} \!\!\right)$
\\
$\rho_{[1-10]}$ & 1 & -1 & $\!\!\! \left( \!\!\renewcommand{\arraycolsep}{2.pt}\renewcommand{\arraystretch}{0.8} \begin{array}{cc}-1&0\\0&1\end{array} \!\!\right)$  & $\!\!\! \left( \!\!\renewcommand{\arraycolsep}{2.pt}\renewcommand{\arraystretch}{0.8} \begin{array}{ccc}0&-1&0 \\-1&0&0 \\ 0&0&-1 \end{array} \!\!\right)$  & $\!\!\! \left( \!\!\renewcommand{\arraycolsep}{2.pt}\renewcommand{\arraystretch}{0.8} \begin{array}{ccc}0&1&0 \\1&0&0 \\ 0&0&1 \end{array} \!\!\right)$  & $\!\!\! \left( \!\!\renewcommand{\arraycolsep}{2.pt}\renewcommand{\arraystretch}{0.8}\begin{array}{cc}0&-1\\ 1&0\end{array} \!\!\right)$
\\
TR    &   1       &  1     & $\!\!\! \left( \!\!\renewcommand{\arraycolsep}{2.pt}\renewcommand{\arraystretch}{0.8} \begin{array}{cc}1&0\\0&1\end{array} \!\!\right)$  & $\!\!\! \left( \!\!\renewcommand{\arraycolsep}{2.pt}\renewcommand{\arraystretch}{0.8}\begin{array}{ccc}-1&0&0 \\0&-1&0 \\ 0&0&-1 \end{array} \!\!\right)$  & $\!\!\! \left( \!\!\renewcommand{\arraycolsep}{2.pt}\renewcommand{\arraystretch}{0.8} \begin{array}{ccc}-1&0&0 \\0&-1&0 \\ 0&0&-1 \end{array} \!\!\right)$  & $\!\!\! \left( \!\!\renewcommand{\arraycolsep}{2.pt}\renewcommand{\arraystretch}{0.8} \begin{array}{cc}0&-i\\i&0\end{array} \!\!\right)$
\\
\hline\hline
\end{tabular}
\end{table}

To avoid ambiguity and for consistency, here we provide explicitly the matrices  we use (i.e., our convention) for all symmetry operations upon the symmetrized states. See Tables~\ref{tab:symmetry_matrices_001}-\ref{tab:symmetry_matrices_110}. The physical conclusions will not depend on the specific convention (and the resulting coefficient phases) we choose as long as it is consistently carried out. We find the following most general symmetry-allowed mixtures without SOC, compatible with the forms of Eqs.~(\ref{eq:S})-(\ref{eq:T+-}).  For $\hat{\mathbf{d}}\|[001]$ we find,
\begin{widetext}
\begin{eqnarray} \label{eq:001_Td_mix}
[001]\;: \; \alpha\uparrow &\Rightarrow& \alpha_{A_1\uparrow} +\delta^i_0\beta_{A_1\uparrow}+ \sum_{(\gamma, t)=(\alpha, c), (\beta, i)}(i\delta^t_1 \gamma_{A_2\uparrow} + i\delta^t_2 \gamma_{E^{I}_z\uparrow} + \delta^t_3 \gamma_{E^{I\!I}_z\uparrow} + \delta^t_4 \gamma_{T_{1z}\uparrow} + i\delta^t_5 \gamma_{T_{2z}\uparrow}),
\nonumber
\\
\beta \downarrow &\Rightarrow& \beta_{A_1\downarrow} +\delta^i_0\alpha_{A_1\downarrow}+
  \sum_{(\gamma, t)=(\beta, c), (\alpha, i)}(i\delta^t_1 \gamma_{A_2\downarrow} + i\delta^t_2 \gamma_{E^{I}_z\downarrow} + \delta^t_3 \gamma_{E^{I\!I}_z\downarrow} - \delta^t_4 \gamma_{T_{1z}\downarrow} - i\delta^t_5 \gamma_{T_{2z}\downarrow} ),
\nonumber
\\
\alpha \downarrow &\Rightarrow& \alpha_{A_1\downarrow} +\delta^i_0\beta_{A_1\downarrow}+ \sum_{(\gamma, t)=(\alpha, c), (\beta, i)}(
- i\delta^t_1 \gamma_{A_2\downarrow} - i\delta^t_2 \gamma_{E^{I}_z\downarrow} + \delta^t_3 \gamma_{E^{I\!I}_z\downarrow} -\delta^t_4 \gamma_{T_{1z}\downarrow} + i\delta^t_5 \gamma_{T_{2z}\downarrow} ),
\nonumber
\\
\beta \uparrow &\Rightarrow& \beta_{A_1\uparrow} +\delta^i_0\alpha_{A_1\uparrow}+ \sum_{(\gamma, t)=(\beta, c), (\alpha, i)}
(- i\delta^t_1 \gamma_{A_2\uparrow} - i\delta^t_2 \gamma_{E^{I}_z\uparrow} + \delta^t_3 \gamma_{E^{I\!I}_z\uparrow} + \delta^t_4 \gamma_{T_{1z}\uparrow} - i\delta^t_5 \gamma_{T_{2z}\uparrow} ),
\end{eqnarray}
\end{widetext}
the first of which is Eq.~(\ref{eq:001_Td_mix-alphaUp}) of the main text. The spins  are along the $z$ direction, and the small coefficients $\delta_i$'s (derived later in Appendix~\ref{app:bbb}) are real, and superscripts $c$ and $i$ denote covalence and ionic, respectively. The $\delta$'s have no relation between each other for different singlet and triplet functions unless they are connected by symmetry in a degenerate IR for a given alignment (i.e., $T_\pm$ in [001] or [111] alignment). Specifically, we note that $\delta^i_{0}=0$ for $T_0$ or $T_\pm$ due to Pauli exclusion. Regarding the restriction imposed by the symmetry, we note in particular the $C_{2z}$ symmetry operation which keeps the atom site and spin orientation. As a result, each mixed component should transform back to itself under $C_{2z}$.  Another constraint is that both $\sigma_{[1\pm 10]}$ and TR connect the same pair of wave functions (i.e., $\alpha \uparrow\leftrightarrow \alpha\downarrow$, $\beta \uparrow\leftrightarrow \beta\downarrow$) up to a global phase difference. This constraint sets the phase of the coefficient for each mixed term, as well as the linear combination of degenerate mixed components.

For $\hat{\mathbf{d}}\|[111]$, we find that the first restriction (by simultaneously satisfying $C^+_{3[111]}$ and $C^-_{3[111]}$) eliminates the possibility of mixing in $E$ IR and two of the three combinations belonging to $T_1$ or $T_2$ IR.
\begin{widetext}
\begin{eqnarray}\label{eq:111_Td_mix}
[111]\;: \;  \alpha \uparrow &\Rightarrow& \alpha_{A_1\uparrow} +\delta^i_0\beta_{A_1\uparrow} +
\sum_{(\gamma, t)=(\alpha, c), (\beta, i)} (i\delta^t_1 \gamma_{A_2\uparrow} +\delta^t_2 \gamma_{T_{1x}+T_{1y}+T_{1z}\uparrow}+i \delta^t_3 \gamma_{T_{2x}+T_{2y}+T_{2z}\uparrow} ),
\nonumber
\\
\alpha \downarrow &\Rightarrow& \alpha_{A_1\downarrow} +\delta^i_0\beta_{A_1\downarrow}+
\sum_{(\gamma, t)=(\alpha, c), (\beta, i)}(- i\delta^t_1 \gamma_{A_2\downarrow} -\delta^t_2 \gamma_{T_{1x}+T_{1y}+T_{1z}\downarrow}+i \delta^t_3 \gamma_{T_{2x}+T_{2y}+T_{2z}\downarrow} ),
\nonumber\\
\alpha &\leftrightarrow& \beta
\end{eqnarray}
\end{widetext}
where spins are along the [111] direction, and the $\delta$'s have no relations between different $\hat{\mathbf{d}}$ alignments. Each $\delta_i$ for $\alpha$ and $\beta$ needs not be to the same for the $[111]n$ case as no symmetry operation in $C_{3v}$ connects atoms $\alpha$ and $\beta$, while it is the same for $[111]i$.   $\delta^i_{0}=0$ for $T_0$ or $T_\pm$.

Finally, for $\hat{\mathbf{d}}\|[110]$, all IR components are allowed to be mixed in, since there is no `self-projecting' operation as in the $\hat{\mathbf{d}}\|$ [001] or [111] case. The $\sigma_{[1-10]}$ and TR symmetry still dictates the linear combination of mutually connected components ($T_{1x}$ and $T_{1y}$; $T_{2x}$ and $T_{2y}$).
\begin{widetext}
\begin{eqnarray}\label{eq:110_Td_mix}
[110]\;: \; \alpha\uparrow &\Rightarrow& \alpha_{A_1\uparrow} +\delta^i_0\beta_{A_1\uparrow}+ \sum_{(\gamma, t)=(\alpha, c), (\beta, i)}
(i\delta^t_1 \gamma_{A_2\uparrow} + i\delta^t_2 \gamma_{E^{I}_z\uparrow} + \delta^t_3 \gamma_{E^{I\!I}_z\uparrow} + \delta^t_4 \gamma_{T_{1z}\uparrow} + i\delta^t_5 \gamma_{T_{2z}\uparrow}+ \delta^t_6 \gamma_{T_{1x}+T_{1y}\uparrow}
 \nonumber\\
 &&\quad\quad+ i\delta^t_7 \gamma_{T_{1x}-T_{1y}\uparrow}+ i\delta^t_8 \gamma_{T_{2x}+T_{2y}\uparrow} + \delta^t_9 \gamma_{T_{2x}-T_{2y}\uparrow} ),
 \nonumber
\\
 \beta\downarrow &\Rightarrow& \beta_{A_1\downarrow}+\delta^i_0\alpha_{A_1\downarrow} + \sum_{(\gamma, t)=(\beta, c), (\alpha, i)}( i\delta^t_1 \gamma_{A_2\downarrow} + i\delta^t_2 \gamma_{E^{I}_z\downarrow} + \delta^t_3 \gamma_{E^{I\!I}_z\downarrow} + \delta^t_4 \gamma_{T_{1z}\downarrow} + i\delta^t_5 \gamma_{T_{2z}\downarrow}- \delta^t_6 \gamma_{T_{1x}+T_{1y}\downarrow}
 \nonumber\\
 &&\quad\quad- i\delta^t_7 \gamma_{T_{1x}-T_{1y}\downarrow}- i\delta^t_8 \gamma_{T_{2x}+T_{2y}\downarrow} - \delta^t_9 \gamma_{T_{2x}-T_{2y}\downarrow} ),
 \nonumber
\\
  \alpha\downarrow &\Rightarrow& \alpha_{A_1\downarrow}  +\delta^i_0\beta_{A_1\downarrow}+\sum_{(\gamma, t)=(\alpha, c), (\beta, i)}
  (- i\delta^t_1 \gamma_{A_2\downarrow} - i\delta^t_2 \gamma_{E^{I}_z\downarrow} + \delta^t_3 \gamma_{E^{I\!I}_z\downarrow} + \delta^t_4 \gamma_{T_{1z}\downarrow} + i\delta^t_5 \gamma_{T_{2z}\downarrow}- \delta^t_6 \gamma_{T_{1x}+T_{1y}\downarrow}
 \nonumber\\
 &&\quad\quad+ i\delta^t_7 \gamma_{T_{1x}-T_{1y}\downarrow}+ i\delta^t_8 \gamma_{T_{2x}+T_{2y}\downarrow} - \delta^t_9 \gamma_{T_{2x}-T_{2y}\downarrow} ),
 \nonumber
\\
 \beta\uparrow &\Rightarrow& \beta_{A_1\uparrow} +\delta^i_0\alpha_{A_1\uparrow} +
 \sum_{(\gamma, t)=(\beta, c), (\alpha, i)}(- i\delta^t_1 \gamma_{A_2\uparrow} - i\delta^t_2 \gamma_{E^{I}_z\uparrow} + \delta^t_3 \gamma_{E^{I\!I}_z\uparrow} - \delta^t_4 \gamma_{T_{1z}\uparrow} + i\delta^t_5 \gamma_{T_{2z}\uparrow}+ \delta^t_6 \gamma_{T_{1x}+T_{1y}\uparrow}
 \nonumber\\
 &&\quad\quad- i\delta^t_7 \gamma_{T_{1x}-T_{1y}\uparrow}- i\delta^t_8 \gamma_{T_{2x}+T_{2y}\uparrow} + \delta^t_9 \gamma_{T_{2x}-T_{2y}\uparrow} ),
\end{eqnarray}
\end{widetext}
where  spins are along the [110] direction. $\delta^i_{0}=0$ for $T_0$ or $T_\pm$.

\section{obtaining mixture coefficients $\delta$'s}\label{app:bbb}

The mixture coefficients $\delta_i$ can be obtained perturbatively. The general procedure for the non-polar mixture is shown in the following. Take the component to be mixed as $X$ IR, which can be any of those in Eqs.~(\ref{eq:001_Td_mix})-(\ref{eq:110_Td_mix}). Together with the unperturbed (denoted by superscript `0') basis states
\begin{eqnarray}
S(T_0)^0: \!\frac{ (1-P_{12})} {2\!\sqrt{1 \!\pm \!\chi_{A_1}^2\!}\!}
[\alpha_{A_1 \!\uparrow}\!(\mathbf{r}_1\!) \beta_{A_1\!\downarrow}\!(\mathbf{r}_2\!)\!\mp \!\alpha_{A_1\!\downarrow} \!(\mathbf{r}_1\!) \beta_{A_1 \!\uparrow}\!(\mathbf{r}_2\!)],
\end{eqnarray}
where $\chi_{A_1 }=\langle \alpha_{A_1}| \beta_{A_1} \rangle$, the generic $S(T_0)$ component with  $A_1$ and $X$ electrons on different atoms has the form
\begin{widetext}
\begin{eqnarray}
\!\!\!S(T_0)': \; \frac{ (1-P_{12})} {\!2\!\sqrt{2(1 \!\pm\! \chi_{A_1 X}^2\!\pm\! \chi_{A_1} \chi_{X})}}
[e^{ia} \!\alpha_{A_1 \!\uparrow}(\mathbf{r}_1\!) \beta_{X\!\downarrow}(\mathbf{r}_2\!) \!+ \!e^{ib} \alpha_{X\! \uparrow}(\mathbf{r}_1\!) \beta_{A_1\!\downarrow}(\mathbf{r}_2\!) \!\mp\! e^{ic} \alpha_{A_1\! \downarrow}(\mathbf{r}_1\!) \beta_{X\!\uparrow}(\mathbf{r}_2\!) \!\mp\! e^{id} \alpha_{X \!\downarrow}(\mathbf{r}_1\!) \beta_{A_1\!\uparrow}(\mathbf{r}_2\!)],\;
\end{eqnarray}
\end{widetext}
where $a, b, c, d$ are determined by symmetry as shown in Eqs.~(\ref{eq:001_Td_mix})-(\ref{eq:110_Td_mix}), $\chi_{A_1 X}=\langle \alpha_{A_1}| \beta_X \rangle$. These coefficients will cancel out due to the same symmetry operations of the two-donor system when one computes $\langle S(T_0)'| H_{\rm int-d}| S(T_0)^0\rangle$ where $H_{\rm int-d}$ is always identity under all symmetry operations. The mixtures for $T_\pm$ states are exactly the same as those for the $T_0$ state if one neglects the small anisotropic SOC effect, which accounts for the energy splitting between the triplet states (the nondegeneracy is shown in Table~\ref{tab:IR} of the main text by symmetry). Between basis states that are not exactly orthogonal, like in our cases due to the finite overlap, the leading-order approximation \cite{Herring_book66} is to replace $H_{\rm int-d}$ with $H_{ee}-\mathcal{E}_{S(T_0)^0}$, where $\mathcal{E}_{S(T_0)^0}=\langle S(T_0)^0| H_{ee}| S(T_0)^0\rangle$ and $H_{ee}$ is the total Hamiltonian for the two-electron system [Eq.~(\ref{eq:H_ee})].
The mixture coefficient of $X$ component follows $ \langle S(T_0)'| H_{ee}- \mathcal{E}_{S(T_0)^0}| S(T_0)^0\rangle/(\mathcal{E}_{S(T_0)}-\mathcal{E}_{S(T_0)'})$, and straightforwardly we have
\begin{widetext}
\begin{eqnarray}
\mathcal{E}_{S(T_0)^0}&=& \frac{ 1} {1 \pm \chi_{A_1}^2} \langle \alpha_{A_1 } \beta_{A_1}| H_{ee}| \alpha_{A_1 } \beta_{A_1} \pm \beta_{A_1} \alpha_{A_1 } \rangle
\nonumber
\\
&=&  \frac{ 1} {1 \pm \chi_{A_1}^2}
 [2\mathcal{E}_{A_1} + \langle \alpha_{A_1 }\beta_{A_1}|\frac{e^2}{r_{12}} | \alpha_{A_1 } \beta_{A_1}\rangle + \frac{e^2}{r_{\alpha\beta}} -2\langle \alpha_{A_1 }|\frac{e^2}{r_{1\beta}} | \alpha_{A_1 } \rangle
\nonumber\\
&&\pm (2 \chi^2_{A_1} \mathcal{E}_{A_1} + \langle \alpha_{A_1 } \beta_{A_1}|\frac{e^2}{r_{12}} | \beta_{A_1 } \alpha_{A_1}\rangle + \chi^2_{A_1} \frac{e^2}{r_{\alpha\beta}} - 2  \chi_{A_1}\langle \alpha_{A_1 } |\frac{e^2}{r_{1\alpha}} | \beta_{A_1 } \rangle)]
\nonumber\\
&\approx& 2\mathcal{E}_{A_1} + \mathcal{O}(e^{-2 r_{\alpha\beta}/a^L_B})
\label{eq:energy_0}
\\
\mathcal{E}_{S(T_0)'}&=& \frac{ 1} {1 \pm \chi_{A_1 X}^2\pm \chi_{A_1} \chi_{X}} \langle e^{ia}\alpha_{A_1 }\beta_{X}| H_{ee}| e^{ia} \alpha_{A_1 } \beta_{X} + e^{ib} \alpha_{X } \beta_{A_1} \pm e^{ic} \beta_{X}\alpha_{A_1 }  \pm e^{id} \beta_{A_1} \alpha_{X} \rangle
\nonumber
\\
&=& \frac{ 1} {1 \pm \chi_{A_1X}^2\pm \chi_{A_1} \chi_{X}}
\{ \mathcal{E}_{A_1}+\mathcal{E}_{X} + \langle \alpha_{A_1 }\beta_{X}|\frac{e^2}{r_{12}} | \alpha_{A_1 } \beta_{X}\rangle + \frac{e^2}{r_{\alpha\beta}} -\langle \alpha_{A_1 }|\frac{e^2}{r_{1\beta}} | \alpha_{A_1 } \rangle -\langle \beta_{X} |\frac{e^2}{r_{2\alpha}} | \beta_{X } \rangle
\nonumber\\
&&+ e^{i(b-a)}\langle \alpha_{A_1 }\beta_{X}|\frac{e^2}{r_{12}} | \alpha_{X } \beta_{A_1}\rangle
\pm e^{i(c-a)}( \langle \alpha_{A_1 } \beta_{X}|\frac{e^2}{r_{12}} | \beta_{X } \alpha_{A_1}\rangle + \chi^2_{A_1X} \frac{e^2}{r_{\alpha\beta}} - 2 \chi_{A_1X} \langle \alpha_{A_1 } |\frac{e^2}{r_{1\beta}} | \beta_{X } \rangle
\nonumber\\
&&
  + 2 \chi^2_{A_1X} \mathcal{E}_{A_1} - 2 \chi_{A_1X} \langle \alpha_{A_1 } |\frac{e^2}{r_{1\beta}} | \beta_{X } \rangle)
 \pm e^{i(d-a)}[ \langle \alpha_{A_1 } \beta_{X}|\frac{e^2}{r_{12}} | \beta_{A_1 } \alpha_{X} \rangle + \chi_{A_1} \chi_{X}  \frac{e^2}{r_{\alpha\beta}}
 \nonumber\\
 &&+
  \chi_{A_1}\chi_{X} (\mathcal{E}_{A_1} +\mathcal{E}_{X}) - \chi_{X} \langle \alpha_{A_1 } |\frac{e^2}{r_{1\beta}} | \beta_{A_1 } \rangle- \chi_{A_1} \langle \beta_{X } |\frac{e^2}{r_{2\beta}} | \alpha_{X } \rangle
]\}
\nonumber\\
&\approx& \mathcal{E}_{A_1}+\mathcal{E}_{X} + \mathcal{O}(e^{-2r_{\alpha\beta}/a^X_B})
\label{eq:energy_1}
\\
\langle S(T_0)'| S(T_0)^0\rangle &=&
\frac{ \sqrt{2 }e^{-ia}} {\sqrt{ (1\pm\chi_{A_1}^2) (1\pm\chi_{A_1X}^2\pm \chi_{A_1} \chi_{X})}} \langle \alpha_{A_1 }\beta_{X}|  \alpha_{A_1 } \beta_{A_1} \pm \beta_{A_1} \alpha_{A_1 } \rangle
\nonumber\\
&=& \pm\frac{ \sqrt{2 }e^{-ia}} {\sqrt{ (1\pm\chi_{A_1}^2) (1\pm\chi_{A_1X}^2\pm \chi_{A_1} \chi_{X})}} \chi_{A_1}\chi_{A_1X}
\label{eq:overlap_two-electron}
\\
\langle S(T_0)'| H_{ee}| S(T_0)^0\rangle &=&
\frac{ \sqrt{2 }e^{-ia}} {\sqrt{ (1\pm\chi_{A_1}^2) (1\pm\chi_{A_1X}^2\pm \chi_{A_1} \chi_{X})}} \langle \alpha_{A_1 } \beta_{X}| H| \alpha_{A_1 } \beta_{A_1} \pm \beta_{A_1} \alpha_{A_1 } \rangle
\nonumber\\
&=&\frac{ \sqrt{2 }e^{-ia}} {\sqrt{ (1\pm\chi_{A_1}^2) (1\pm\chi_{A_1X}^2\pm \chi_{A_1} \chi_{X})}}
 [\langle \alpha_{A_1 }\beta_{X}|\frac{e^2}{r_{12}} | \alpha_{A_1 } \beta_{A_1}\rangle  - \langle \beta_{X}|\frac{e^2}{r_{2\alpha}} | \beta_{A_1 } \rangle
\nonumber\\
&&\pm (\langle \alpha_{A_1 }\beta_{X}|\frac{e^2}{r_{12}} | \beta_{A_1 } \alpha_{A_1}\rangle +   \chi_{A_1}\chi_{A_1 X} \frac{e^2}{r_{\alpha\beta}} + 3 \chi_{A_1}\chi_{A_1 X} \mathcal{E}_{A_1} -  \chi_{A_1}\langle \beta_{X } |\frac{e^2}{r_{2\beta}} | \alpha_{A_1 } \rangle)
]
\nonumber\\
&\sim& \mathcal{O}(e^{- r_{\alpha\beta}(1/a^L_B +1/a^X_B)})
\label{eq:two-electron_couple_H}
\end{eqnarray}
\end{widetext}
where $\mathbf{r}_{1,2}$ is omitted and the first (second) wavefunction in the state belongs to the first (second) electron, and the dielectric constant $\epsilon$ is also omitted above. In Eq.~(\ref{eq:energy_0}), the second to fourth terms are the direct Coulomb energy of two neutral atoms and sum to $\sim e^{-2 r_{\alpha\beta}/a^L_B}$ \cite{Heitler_London_ZP27}, where unlike the hydrogen atoms the unperturbed $S(T_0)^0$ of Si donors is formed by the $1s$ state whose decay length are $a^{L(S)}_B$ along the larger(smaller) effective mass direction. In Eq.~(\ref{eq:energy_1}), again the third to seventh terms sum up to $\sim e^{-2 r_{\alpha\beta}/a^X_B}$  where $a^X_B$ is about the longest decay length in the $X$'s envelope. In Eq.~(\ref{eq:two-electron_couple_H}), the first two terms are the coupling of $\beta_{A_1}$ and $\beta_{X}$ by a neutral $\alpha$ atom and largely cancel out.  Later we will see that the mixture coefficients due to the direct Coulomb interaction cancel out when calculating the relaxation matrix elements, leaving the physical effects  induced  only by exchange interaction. Quantitatively, the net effect of the inter-donor perturbation is to couple excited donor states with weights the order of $J(\alpha_{A_1} \beta_{A_1};\alpha_{A_1}\beta_X)/(\mathcal{E}_{A_1}-\mathcal{E}_{X})$. The numerator is similar to the $S-T$ splitting involving the given envelopes in $A_1$ (i.e., $1s$) and $X$ (similar arguments in $J$ estimation in \cite{Mariya_PR60}).  Furthermore, for relaxation between triplet states, the mixture coefficients cancel out except for the effect of anisotropic SOC not shown above. We can also examine the mixture with polar excited states following a similar procedure. Again, the important part is the difference between mixture coefficients in the singlet and triplet states.

\section{Details on perturbative expansion of Eqs.~(\ref{eq:S-T0-001})-(\ref{eq:T-barT-110})} \label{app:ccc}

Using the physical principles discussed in the main text for identifying the strongest coupling strengths among all perturbations, we obtain the essential perturbation expansions of the relaxation matrix elements derived in Eqs.~(\ref{eq:S-T0-001})-(\ref{eq:T-barT-110})  one by one.

We start with Eq.~(\ref{eq:S-T0-001}) of $\langle T_0| \epsilon_{xx}-\epsilon_{yy} | S \rangle_{[001]}$. As discussed, inter-donor interaction, $H_{\rm int-d}$, or overlap has to be involved to have finite relaxation. In the first term, there are two ways to have perturbation made by $H_{\rm int-d}$:  the mixture of non-polar (covalent) excited states, or the mixture of the polar (ionic) excited states. For the non-polar ones, the one-electron mixture is on the same donor [see Eqs.~(\ref{eq:001_Td_mix})-(\ref{eq:110_Td_mix})] and the coefficient is proportional to $e^{-i r_{\alpha\beta}(1/a^L_B +1/a^X_B)}$. The polar ones  require another overlap factor. In the second term of Eq.~(\ref{eq:S-T0-001}), one factor utilizes the polar mixture with coefficient  $\propto e^{-r_{\alpha\beta}/a^X_B}$ while the other factor goes with overlap $\propto e^{-r_{\alpha\beta}/a^L_B}$, and is comparable to the first term (the double overlap contribution mostly cancels out due to TR symmetry).

Next we study the first term due to its non-polar coupling by $H_{\rm int-d}$. We find out that the dominant term comes with donor impurity  SOC $\sim L_z\sigma_z$. $L_z$ belongs to the $T_{1z}$ IR of the $T_d$ group ($L_z \sim T_{1z}$). It only couples the same $\Delta$ band strongly. Then the $H_{\rm ep}$ has to couple the same $\Delta$ band (in addition to the same envelope state) as a result. We find this is possible, although $\epsilon_{xx}-\epsilon_{yy}\sim \Delta^z_2$ in the $z$ valleys, $\epsilon_{xx(yy)}\sim \Delta^{x(y)}_{1}$, and $\epsilon_{yy(xx)}=\frac{1}{2}[(\epsilon_{yy(xx)}+\epsilon_{zz})+(\epsilon_{yy(xx)}-\epsilon_{zz})] \sim \Delta^{x(y)}_1 + \Delta^{x(y)}_2$ in the $x(y)$ valleys. $\epsilon_{xx}-\epsilon_{yy}$ belongs to the $E^{I}_z$ IR of the $T_d$ group, to which one of the donor state configuration out of the $\Delta_1$ band also belongs. So $\epsilon_{xx}-\epsilon_{yy}$ couples $\Delta_1-1s-A_1$ to $\Delta_1-1s-E^I_z$, which in turn is coupled by $L_z$ to $\Delta_1-1s-T_{2z}$ ($E^I_z\times T_{1z} =T_{2z}$), which is finally coupled back to $\Delta_1-1s-A_1$ by $H_{\rm int-d}$ [allowed in Eq.~(\ref{eq:001_Td_mix})].  For the polar mixture contribution, we need to keep everything except replacing the $T_{2z}$ mixture in the same donor to all the possible mixtures $X$'s in the other donor by $H_{\rm int-d}$ each multiplying an overlap $\langle \alpha_{T_{2z}}|\beta_{X}\rangle$. For the second term of Eq.~(\ref{eq:S-T0-001}), we can replace the inter-donor mixture by a simple overlap, multiplied by a polar mixture with the $T_{2z}$ IR.

As the combined interaction of $H_{\rm ep}$ with diagonal strain elements (which give rise to deformation potential within the $\Delta_1$ band) and SOC is used frequently, we show their combined $T_d$ symmetry explicitly. $T_1\times E = T_1+T_2$,
\begin{eqnarray}\label{eq:sigma_xy_epsilon_E_Td}
T_1&:&  \{(\epsilon_{yy}+\epsilon_{zz}-2 \epsilon_{xx})\sigma_x,  (\epsilon_{xx}+\epsilon_{zz}-2 \epsilon_{yy})\sigma_y,
 \nonumber\\
 &&\;(\epsilon_{xx}+\epsilon_{yy}-2 \epsilon_{zz})\sigma_z\},
\nonumber\\
T_2&:&  \{(\epsilon_{yy}-\epsilon_{zz})\sigma_x,  (\epsilon_{zz}-\epsilon_{xx})\sigma_y,  (\epsilon_{xx}-\epsilon_{yy})\sigma_z\}.\quad
\end{eqnarray}

The most important one of the  perturbation expressions follows as Eq.~(\ref{eq:S-T0-001-1}) of the main text, plus the term reversing perturbation ordering. From Eqs.~(\ref{eq:energy_0})-(\ref{eq:overlap_two-electron}), we know that the two coefficients are differed by parts due to the exchange interaction integrals. By reversing the matrix element ordering in the numerator, it essentially takes the TR operation. The key is that $L_i$ is odd under TR, while $\epsilon_{ij}$ and $H_{\rm int-d}$ are even under TR. These parities yield a minus sign between these two orderings, and as a result the direct Coulomb integrals of the inter-donor mixture in $S$ and $T_0$ cancel out. This cancellation makes physical sense, as the singlet and triplets respond the same way to the direct Coulomb part of the interaction. The otherwise cancellation between two same $S$ (or $T$) states can also be generally reached by TR,
\begin{eqnarray}\label{eq:cancel_by_TR}
\langle \psi^{S}_{\alpha}|\epsilon_{ij} L_k| \psi^S_{\alpha}\rangle
=-\langle \psi^{S}_{\alpha}|\epsilon_{ij} L_k| \psi^S_{\alpha}\rangle,
\end{eqnarray}
for any $i,j,k$ components.

The other two orderings involve $T_{1z}$ states which cannot be formed out of $s$ envelopes but  only with $d_{\pm1}$ envelopes,
\begin{widetext}
\begin{eqnarray}
\langle T_0| \epsilon_{xx}-\epsilon_{yy} | S \rangle_{[001]}^{(2)}
&\propto& \frac{\langle \Delta_1\!,1s\!,A_1|L_z|\Delta_1\!,3d_{\pm1}\!,T_{1z}\rangle \langle \Delta_1\!,3d_{\pm1}\!,T_{1z}| \overline{H}_{\rm int-d}|\Delta_1\!,1s\!,E^I_z\rangle \langle \Delta_1\!,1s\!,E^I_z| \epsilon_{xx}-\epsilon_{yy} |\Delta_1\!,1s\!,A_1\rangle}
{[\mathcal{E}(\Delta_1,1s,A_1)- \mathcal{E}(\Delta_1,3d_{\pm1},T_1)] [\mathcal{E}(\Delta_1,1s,A_1)- \mathcal{E}(\Delta_1,1s,E^I)]}.
\nonumber\\
&&+ \textrm{terms reversing ordering}.
\label{eq:S-T0-001-2}
\end{eqnarray}
\end{widetext}
This term is similar in magnitude to those in Eq.~(\ref{eq:S-T0-001-1}) involving $3d$ intermediate states.
\begin{widetext}
\begin{eqnarray}
\langle T_0| \epsilon_{xx}-\epsilon_{yy} | S \rangle_{[001]}^{(3)}
&\propto& \frac{\langle \Delta_1\!,1s\!,A_1| \overline{H}_{\rm int-d}|\Delta_1\!,3d_{\pm1}\!,T^z_2\rangle \langle \Delta_1\!,3d_{\pm1}\!,T_{2z}| \epsilon_{xx}-\epsilon_{yy} |\Delta_1\!,3d_{\pm1}\!,T_{1z}\rangle \langle \Delta_1\!,3d_{\pm1}\!,T_{1z}| L_z |\Delta_1\!,1s\!,A_1\rangle}
{[\mathcal{E}(\Delta_1,1s,A_1)- \mathcal{E}(\Delta_1,3d_{\pm1},T_1)] [\mathcal{E}(\Delta_1,1s,A_1)- \mathcal{E}(\Delta_1,3d_{\pm1},T_2)]}
\nonumber\\
&&+ \textrm{terms reversing ordering}.
\label{eq:S-T0-001-3}
\end{eqnarray}
\end{widetext}
 This term is smaller than Eq.~(\ref{eq:S-T0-001-2}) because of the two relatively large energy denominators instead of one.

For Eq.~(\ref{eq:S-T0-110}) of $\langle T_0|\epsilon_{z''y''} | S \rangle_{[110]}$, the first and third terms dominate by one less SOC. As for the $\hat{\mathbf{d}}\|$[001] case, $L_{z''}=\frac{1}{\sqrt{2}}(L_{x} +L_{y})$ is required in conjunction with $\epsilon_{z''y''}$ to couple $\langle \psi^{T_0}_{\alpha}|...| \psi^{S}_{\alpha(\beta)}\rangle$, due to the $\sigma_{y''}$ reflection symmetry. $\epsilon_{z''y''}\equiv \epsilon_{[110][1-10]}=\epsilon_{xx}-\epsilon_{yy}$. And we have a coupling similar to that for Eq.~(\ref{eq:S-T0-001}). Using Eq.~(\ref{eq:sigma_xy_epsilon_E_Td}), the combined interaction is symmetrized as,
\begin{eqnarray}\label{eq:e-ph-SOC-Td-S-T0-110}
&&\!\!\!\!\!(\epsilon_{xx}\!-\!\epsilon_{yy})(L_{x} \!+\!L_{y})
\!=\!
-\frac{1}{2} [(\epsilon_{yy}\!-\!\epsilon_{zz})L_x\! +\! (\epsilon_{zz}\!-\!\epsilon_{xx})L_y] \quad
\nonumber\\
&&\quad-\frac{1}{2} [(\epsilon_{yy}+\epsilon_{zz}-2\epsilon_{xx})L_x + (\epsilon_{xx}+\epsilon_{zz}-2\epsilon_{yy})L_y]
\nonumber\\
&& \quad \sim(T^x_2+T^y_2) + (T^x_1+T^y_1).
\end{eqnarray}
With the $T^x_2+T^y_2$ part,  we can follow Eqs.~(15), (\ref{eq:S-T0-001-2}), and (\ref{eq:S-T0-001-3}) and replace $(\epsilon_{xx}-\epsilon_{yy})L_z$ with $(\epsilon_{yy}-\epsilon_{zz})L_x$ and $(\epsilon_{zz}-\epsilon_{xx})L_y$. With the $T^x_1+T^y_1$ part, the perturbation expansion is similar except that there is at least one denominator energy $\sim 40$ meV, since $T_1$ cannot be formed from $1s$ envelopes.

For Eq.~(\ref{eq:S-T+-001}) of $\langle T_+| \epsilon_{xz}-i\epsilon_{yz} | S \rangle_{[001]}$, since $\epsilon_{xz}-i\epsilon_{yz}$ does not partially belong to the $\Delta_1$ in any valley, it has to go with interband coupling, which cannot be brought back to the $\Delta_1$ band state considerably by donor SOC or $H_{\rm int-d}$. As noted in the main text, the host SOC leads to an E-Y cancellation. In all cases, the contributing terms are substantially smaller than Eq.~(\ref{eq:S-T0-001-1}).

For Eq.~(\ref{eq:S-T+-111n}) of $\langle T_+| \epsilon_{x'z'}+i\epsilon_{y'z'} | S \rangle_{[111]n}$, the first (third) term cannot be connected to the second (fourth) term  due to the lack of $\alpha \leftrightarrow \beta$ symmetry operations (they are constructive in the $\hat{\mathbf{d}}\|$[001] case and destructive in the $\hat{\mathbf{d}}\|[111]i$ case). Now we go into more details about them.  $\langle \uparrow|
L_{x'}\sigma_x +L_{y'}\sigma_y|\downarrow\rangle = L_{x'}-i L_{y'} =\sqrt{\frac{2}{3}}[e^{i\pi/6} L_x-iL_y -e^{-i\pi/6}  L_z]$ is necessary. $\epsilon_{x'z'}+i\epsilon_{y'z'} = \sqrt{2}/3[e^{-i\pi/6} \epsilon_{xx} +i\epsilon_{yy} -e^{i\pi/6} \epsilon_{zz} +e^{i\pi/6} \epsilon_{xy}-i \epsilon_{xz} -e^{-i\pi/6}\epsilon_{yz}]$.  Since the e-ph part contains both the diagonal and off-diagonal strain tensor elements, we have respectively the perturbation within the $\Delta_1$ band (using diagonal $\epsilon_{ii}$ with donor SOC) and via the $\Delta_5$ band (using $\epsilon_{ij},i\neq j$ and host SOC), the latter of which is dropped due to E-Y cancellation.

We study for the diagonal e-ph part. The coefficients in $e^{-i\pi/6} \epsilon_{xx} +i\epsilon_{yy} -e^{i\pi/6} \epsilon_{zz}$ add up to 0, showing there is no $\epsilon_{xx}+\epsilon_{yy}+\epsilon_{zz}\sim A_1$ component. Therefore it contains only $E$ IR components. We know $L_i\sim T_{1i}$. $E\times T_1 = T_1+T_2$. So in the first case the combined e-ph and SOC mixes in some components of $T_1$ or $T_2$. Now we check Eq.~(\ref{eq:111_Td_mix}) to see whether and what $T_{1,2}$ components can be coupled by $H_{\rm int-d}$. Indeed, not all the components but only $T_{1x}+T_{1y}+T_{1z}$ and $T_{2x}+ T_{2y}+ T_{2z}$ are allowed. In order to see whether they are contained in the combined e-ph and SOC, we use Eq.~(\ref{eq:sigma_xy_epsilon_E_Td}) to make symmetrization,
\begin{eqnarray}
&&\!\!(e^{-i\pi/6} \epsilon_{xx} \!+\!i\epsilon_{yy}\! -\!e^{i\pi/6} \epsilon_{zz} )(e^{i\pi/6} L_x\!- \! iL_y \!-\!e^{-i\pi/6}  L_z)
\nonumber\\
&\!\!=& \!\!\frac{\sqrt{3}i}{2}[ (\epsilon_{yy}-\epsilon_{zz})L_x + (\epsilon_{zz}-\epsilon_{xx})L_y +(\epsilon_{xx}-\epsilon_{yy})L_z]\quad
\nonumber\\
&&- \frac{1}{2}[(\epsilon_{yy}+\epsilon_{zz} -2\epsilon_{xx})L_x + (\epsilon_{xx}+\epsilon_{zz} -2\epsilon_{yy})L_y
\nonumber\\
&&\qquad+(\epsilon_{xx}+\epsilon_{yy} - 2\epsilon_{zz})L_z]
,
\end{eqnarray}
and find out that it does contain both $\sum_i T_{1i}$ and $\sum_i T_{2i}$  components. Basically, this is a very similar situation as shown in Eq.~(\ref{eq:e-ph-SOC-Td-S-T0-110}) of the [110] $S-T_0$ case. The major contributing terms in the first (or second) term are identified as follows,
\begin{widetext}
\begin{eqnarray}\label{eq:S-T+-111-perturbation}
&&\langle \psi^{T_+}_{\alpha\uparrow'}|  e^{-i\pi/6} \epsilon_{xx} +i\epsilon_{yy} -e^{i\pi/6} \epsilon_{zz}| \psi^{S}_{\alpha\downarrow'}\rangle_{[111]n}
\nonumber\\
&\propto&
\sum_{\substack  {i=x,y,z; \\n=1s,3d_{\pm1}}}\frac{\sqrt{3}i}{2}\frac{\langle \Delta_1\!,1s\!,A_1| \overline{H}_{\rm int-d}|\Delta_1\!,n\!,T_{2i}\rangle
 \langle \Delta_1\!,n\!,T_{2i}| L_i|\Delta_1\!,1s\!,E^I_i\rangle
 \langle \Delta_1\!,1s\!,E^I_i|(\epsilon_{jj}-\epsilon_{kk}) |\Delta_1\!,1s\!,A_1\rangle}
{[\mathcal{E}(\Delta_1,1s,A_1)- \mathcal{E}(\Delta_1,n,T_2)][\mathcal{E}(\Delta_1,1s,A_1)- \mathcal{E}(\Delta_1,1s,E)]}
\nonumber\\
&&+
\sum_{i=x,y,z;}\frac{\sqrt{3}i}{2}\frac{\langle \Delta_1\!,1s\!,A_1| \overline{H}_{\rm int-d}|\Delta_1\!,3d_{\pm1}\!,T_{2i}\rangle
 \langle \Delta_1\!,3d_{\pm1}\!,T_{2i}| (\epsilon_{jj}-\epsilon_{kk})|\Delta_1\!,3d_{\pm1}\!,T_{1i}\rangle
 \langle \Delta_1\!,3d_{\pm1}\!,T_{1i}|L_i |\Delta_1\!,1s\!,A_1\rangle}
{[\mathcal{E}(\Delta_1,1s,A_1)- \mathcal{E}(\Delta_1,3d_{\pm1},T_2)][\mathcal{E}(\Delta_1,1s,A_1)- \mathcal{E}(\Delta_1,3d_{\pm1},T_1)]}
\nonumber\\
&&+
\sum_{i=x,y,z;}\frac{\sqrt{3}i}{2}\frac{\langle \Delta_1\!,1s\!,A_1| (\epsilon_{jj}-\epsilon_{kk})|\Delta_1\!,1s\!,E^I_i\rangle
 \langle \Delta_1\!,1s\!,E^I_i|\overline{H}_{\rm int-d}|\Delta_1\!,3d_{\pm1}\!,T_{1i}\rangle
 \langle \Delta_1\!,3d_{\pm1}\!,T_{1i}|L_i |\Delta_1\!,1s\!,A_1\rangle}
{[\mathcal{E}(\Delta_1,1s,A_1)- \mathcal{E}(\Delta_1,1s,E)][\mathcal{E}(\Delta_1,1s,A_1)- \mathcal{E}(\Delta_1,3d_{\pm1},T_1)]}
\nonumber\\
&&-
\sum_{i=x,y,z}\frac{1}{2}\frac{\langle \Delta_1\!,1s\!,A_1| \overline{H}_{\rm int-d}|\Delta_1\!,3d_{\pm1}\!,T_{1i}\rangle
 \langle \Delta_1\!,3d_{\pm1}\!,T_{1i}| L_i|\Delta_1\!,1s\!,E^{I\!I}_i\rangle
 \langle \Delta_1\!,1s\!,E^{I\!I}_i|(\epsilon_{jj}+\epsilon_{kk} -2\epsilon_{ii}) |\Delta_1\!,1s\!,A_1\rangle}
{[\mathcal{E}(\Delta_1,1s,A_1)- \mathcal{E}(\Delta_1,3d_{\pm1},T_1)][\mathcal{E}(\Delta_1,1s,A_1)- \mathcal{E}(\Delta_1,1s,E)]}
\nonumber\\
&&-
\sum_{i=x,y,z}\frac{1}{2}\frac{\langle \Delta_1\!,1s\!,A_1| \overline{H}_{\rm int-d}|\Delta_1\!,3d_{\pm1}\!,T_{1i}\rangle
 \langle \Delta_1\!,3d_{\pm1}\!,T_{1i}|(\epsilon_{jj}+\epsilon_{kk} -2\epsilon_{ii})|\Delta_1\!,3d_{\pm1}\!,T_{1i}\rangle
 \langle \Delta_1\!,3d_{\pm1}\!,T_{1i}|L_i |\Delta_1\!,1s\!,A_1\rangle}
{[\mathcal{E}(\Delta_1,1s,A_1)- \mathcal{E}(\Delta_1,3d_{\pm1},T_1)][\mathcal{E}(\Delta_1,1s,A_1)- \mathcal{E}(\Delta_1,3d_{\pm1},T_1)]}
\nonumber\\
&&-
\sum_{i=x,y,z}\frac{1}{2}\frac{\langle \Delta_1\!,1s\!,A_1| (\epsilon_{jj}+\epsilon_{kk} -2\epsilon_{ii})|\Delta_1\!,1s\!,E^{I\!I}_i\rangle
 \langle \Delta_1\!,1s\!,E^{I\!I}_i|\overline{H}_{\rm int-d}|\Delta_1\!,3d_{\pm1}\!,T_{1i}\rangle
 \langle \Delta_1\!,3d_{\pm1}\!,T_{1i}|L_i |\Delta_1\!,1s\!,A_1\rangle}
{[\mathcal{E}(\Delta_1,1s,A_1)- \mathcal{E}(\Delta_1,1s,E)][\mathcal{E}(\Delta_1,1s,A_1)- \mathcal{E}(\Delta_1,3d_{\pm1},T_1)]}
\nonumber\\
&&+ \textrm{terms reversing ordering},
\end{eqnarray}
\end{widetext}
where $i,j,k$ ordered cyclicly. Polar mixture also exist in all four terms of Eq.~(\ref{eq:S-T+-111n}) similarly to that of Eq.~(\ref{eq:S-T0-001}).

For Eq.~(\ref{eq:S-T12-110}) of $\langle T_++T_-| \epsilon_{A_1} | S \rangle_{[110]}$, the first term consists of one spin-independent matrix element,  $\langle \psi^{T}_{\alpha\uparrow}| \epsilon_{A_1}| \psi^{S}_{\alpha\uparrow}\rangle\approx \langle \psi_{A_1}| \epsilon_{A_1}| \psi_{A_1}\rangle=\Xi_d+\Xi_u/3$ (which are the usual dilation and shear deformation potentials \cite{Yu_Cardona_Book}), and one spin-flip overlap $\langle \psi^{T_2}_{\beta\uparrow''} | \psi^{S}_{\beta\downarrow''} \rangle$, which we treat similarly to the manner above:
\begin{widetext}
\begin{eqnarray}
\langle \psi^{T}_{\beta\uparrow''} | \psi^{S}_{\beta\downarrow''} \rangle_{[110]}
&\propto&
\frac{\langle \Delta_1\!,1s\!,A_1| \overline{H}_{\rm int-d}|\Delta_1\!,3d_{\pm1}\!,T_1,\!n\rangle
 \langle \Delta_1\!,3d_{\pm1}\!,T_1| L_z-\frac{i}{\sqrt{2}}(L_x-L_y) |\Delta_1\!,1s\!,A_1\rangle}
{\mathcal{E}(\Delta_1,1s,A_1)- \mathcal{E}(\Delta_1,3d_{\pm1},T_1)}
\nonumber\\
&&+ \textrm{terms reversing ordering},
\end{eqnarray}
\end{widetext}
relatively small due to the $3d$ intermediate state. The second term contains  $\langle \psi^{T}_{\alpha\uparrow''}| \epsilon_{A_1}| \psi^{S}_{\alpha\downarrow''}\rangle$, where $\epsilon_{A_1}$ stands for $\epsilon_{x''x''}\equiv  \epsilon_{zz}$ and  $\epsilon_{z''z''(y''y'')}\equiv \epsilon_{[1\pm 10][1\pm 10]}=(\epsilon_{xx}+\epsilon_{yy}\pm 2\epsilon_{xy})/2$.  We can similarly symmetrize the operator as before. Seeking the dominant $T_2$ components, we take from $\epsilon_{zz}$ or $\epsilon_{xx}+\epsilon_{yy}$ the $E^{I\!I}_z$ part (dropping the $A_1$ part), $\sim \epsilon_{xx}+\epsilon_{yy}-2 \epsilon_{zz}$, and the $L_x-L_y$ part from $L_{x''}-iL_{y''}$ to obtain,
\begin{eqnarray}
&&(\epsilon_{xx}+\epsilon_{yy}-2 \epsilon_{zz})(L_x-L_y)
\nonumber\\
&&\qquad\qquad= (\epsilon_{yy}-\epsilon_{zz})L_x +(\epsilon_{zz}-\epsilon_{xx})L_y ,
\end{eqnarray}
similarly to Eq.~(\ref{eq:e-ph-SOC-Td-S-T0-110}), which  the rest of the steps follow.

For Eq.~(\ref{eq:S-T12-110}) of $\langle T_+-T_-| \epsilon_{x''y''} | S \rangle_{[110]}$, SOC is needed such that $\langle \psi^{T_2}_{\alpha(\beta)}| \epsilon_{x''y''}L_{z''}| \psi^{S}_{\alpha}\rangle$ is nonvanishing (due to $\sigma_{y''}$), so the first and third terms require SOC$^2$, much smaller than the second and fourth terms. For $\langle \psi^{T_2}_{\alpha\uparrow''}| \epsilon_{x''y''}| \psi^{S}_{\alpha\downarrow''}\rangle $, $\epsilon_{x''y''}=\frac{1}{\sqrt{2}} (\epsilon_{xz}-\epsilon_{yz})$. We have to go through the $\Delta_5$ intermediate state and host SOC. The result is suppressed by the E-Y cancellation.

\textit{The relaxation matrix elements between different triplet ($T$) states}. As discussed in the main text [also see Eq.~(\ref{eq:cancel_by_TR})], the relaxation does not vanish only when the mixtures into the two states by inter-donor interaction are not the same. This can be satisfied between $S$ and $T$ states by exchange interaction even the crystal field is completely uniform. Between the $T$ states, by the same reason, the different mixtures have to come from field anisotropy with respect to triplet spin orientations. The same origin leads to the small splitting ($I$, much smaller than the $S$-$T$ splitting, $J$) between different triplet states. We briefly discuss the perturbation expansion from these matrix elements.

For Eq.~(\ref{eq:S-T+-001})  of $\langle T_+| \epsilon_{xz}+i\epsilon_{yz} | T_0 \rangle_{[001]}$, due to the non-diagonal e-ph element, $\epsilon_{xz}+i\epsilon_{yz}$, the host SOC is invoked and E-Y cancellation is switched on [as for Eq.~(\ref{eq:S-T+-001}) of $\langle T_+| \epsilon_{xz}-i\epsilon_{yz} | S \rangle_{[001]}$] making this matrix element  small even among ones for the triplet relaxations.

For Eq.~(\ref{eq:S-T+-111n}) of $\langle T_+| \epsilon_{x'z'}+i\epsilon_{y'z'} | T_0 \rangle_{[111]n}$, the e-ph and SOC couplings follow those of $\langle T_+| \epsilon_{x'z'}+i\epsilon_{y'z'} | S \rangle_{[111]n}$ in Eq.~(\ref{eq:S-T+-111n}), while the exchange part is much suppressed as discussed above. The first (third) and second (fourth) terms become equal to each other in Eq.~(\ref{eq:T0-T+-111i}) of $\langle T_+| \epsilon_{x'z'}\!+\!i\epsilon_{y'z'} | T_0 \rangle_{[111]i}$.

In Eq.~(\ref{eq:T0-T12-110}) of $\langle T_+ + T_-|  \epsilon_{y''z''}  |T_0 \rangle_{[110]}$, the first and third terms scale with SOC$^2$ (without taking the inter-donor SOC effect into account) and are much smaller than the second and fourth terms. $\epsilon_{y''z''}=(\epsilon_{xx}-\epsilon_{yy})/\sqrt{2}\sim E^{I}_z$, and  $L_{x''}-iL_{y''}=L_z-\frac{i}{\sqrt{2}}(L_{x}- L_y)\sim T_1$. $E\times T_1=T_1+T_2$. So the perturbation expansion has a similar form  to that in Eq.~(\ref{eq:S-T+-111-perturbation}).

Finally, in  Eq.~(\ref{eq:T0-T12-110}) of $\langle T_+-T_-|  \epsilon_{x''z''}  |T_0 \rangle_{[110]}$ and (\ref{eq:T-barT-110}) of $\langle T_+ -T_-| \epsilon_{x''y''} | T_++T_- \rangle_{[110]}$, $\epsilon_{x''z''} =\frac{1}{\sqrt{2}}(\epsilon_{xz}+\epsilon_{yz})$ and $\epsilon_{x''y''} =\frac{1}{\sqrt{2}} (\epsilon_{xz}-\epsilon_{yz})$ both invoke E-Y cancellation and suppress the two matrix elements even further, as in Eq.~(\ref{eq:S-T+-001}) of $\langle T_+| \epsilon_{xz}+i\epsilon_{yz} | T_0 \rangle_{[001]}$.


\begin{thebibliography}{99}
\bibitem{footnote_2SOC} The donor SOC was not considered, for example, in Refs.~\cite{Hasegawa_PR60_Roth_PR60_Roth_LincolnLab60} due to the measured small dependence of electron $g$-factor on donor types \cite{Feher_PR59} indicating its weak interband coupling. On the other hand, the strong donor-dependent spin split clearly manifests  the relevance of donor  SOC \cite{Castner_PR67, Song_PRL14}, which leads to comparable or stronger spin mixing ($\sim$ 0.03 meV/12 meV for Si:P) than that induced by host SOC ($\sim$ 4.1 meV/4.3 eV, see, e.g., \cite{Li_PRL11,Song_PRB12}).
\bibitem{footnote_3d} We obtain 7.2\% true $3d$ weight inside the $1s$ donor envelope. This SOC coupling is estimated to be 0.3 $\mu$eV \cite{Sugihara_JPSJ63}. It is partially compensated by the $H_{\rm int-d}$ integral due to larger $3d$ spread. From a upper-limit estimate \cite{Herring_book66}, $J\propto \exp[-d(\frac{1}{a_B}+\frac{1}{3a_B})]$ where $a_B$ is the Bohr radius in Si (before considering a suppressing  power-law  factor), it is sufficient to tell that the exchange integral does not completely compensate the 100 times smaller SOC coupling (and another 3.3 times from the energy denominator) for typical $d/a_B\lesssim 5$.
\end{thebibliography}
\end{document}